\documentclass[twocolumn,showpacs,showkeys,prb]{revtex4}
\newcommand{\lfunc}[1]{\left(#1_L^{(0)}\right|}
\newcommand{\lvec}[1]{\left|#1_L\right)}
\newcommand{\En}{\mathcal E}
\newcommand{\rvec}[1]{\left|#1_R\right)}
\newcommand{\fvec}[1]{\left|#1\right>}
\newcommand{\ffun}[1]{\left<#1\right|}
\newcommand{\fxpv}[1]{\left<#1\right>}
\newcommand{\bfvec}[1]{\left|#1\right)}
\newcommand{\bffun}[1]{\left(#1\right|}

\newcommand{\comm}[1]{\left[#1\right]}

\newcommand{\dmap}{\mathcal T_D}
\newcommand{\bfspace}{\mathcal H_{BF}}

\newcommand{\mcq}{\mathcal A}
\newcommand{\mcs}{\mathcal S}
\newcommand{\nf}{\mathcal N_F}
\newcommand{\nb}{\mathcal N_B}
\newcommand{\tr}{\,{\rm tr}_1\,}

\begin{document}


\title{Analysis of the Strong Coupling Limit of the Richardson Hamiltonian\\
using the Dyson Mapping}


\author{I. Snyman}
\email[E-mail: ]{isnyman@lorentz.leidenuniv.nl}
\affiliation{Institute of Theoretical Physics, University of Stellenbosch,\\
Private Bag X1, Matieland 7602, South Africa.\\
and\\
Instituut-Lorentz for Theoretical Physics, Universiteit Leiden,\\
P.O. Box 9506, NL-2300 RA Leiden, The Netherlands.}
\author{H.B. Geyer}
\email[E-mail: ]{hbg@sun.ac.za}
\affiliation{Institute of Theoretical Physics, University of Stellenbosch,\\
Private Bag X1, Matieland 7602, South Africa.}


\date{\today}

\begin{abstract}
The Richardson Hamiltonian describes superconducting correlations in a metallic
nanograin. We do a perturbative analysis of this and related Hamiltonians, around 
the strong pairing limit, without having to invoke the Bethe Ansatz
solvability. Rather we make use of a boson expansion method known as the Dyson mapping.
Thus we uncover a selection rule that facilitates both time-independent and time-dependent 
perturbation expansions. In principle the model we analise is realised in
a very small metalic grain of a very regular shape. The results we obtain point to
subtleties sometimes neglected when thinking of the superconducting state as a
Bose-Einstein condensate. An appendix contains a general presentation of
time-independent perturbation theory for operators with degenerate spectra, with recursive 
formulas for corrections of arbitrarily high orders.
\end{abstract}

\pacs{71.15.-m,71.10.Ca,74.20.-z}
\keywords{strongly correlated fermions, superconductivity, boson expansion methods}

\maketitle
\section{\label{sect0}Introduction}
Superconduction correlations of electrons in weakly disordered metallic nano-grains\cite{kaa00,vdr01} 
are quite generally described by the Richardson Hamiltonian. This model was first introduced in 
the sixties in the context of nuclear physics\cite{ric63a,ric63b} and is Bethe Ansatz solvable.
This means that the spectrum is known in terms of a set of parameters whose possible values
are determined as the simultaneous roots of a large system of (non-linear) equations that have
to be solved numerically.
 
Most studies of superconduction correlations in nano-grains therefore heavily rely on numerical
solutions of the Bethe Ansatz equations for the Richardson Hamiltonian. What remained lacking
was a simple intuitive picture of the low-energy physics of such grains, that would come from
analytical rather than numerical studies. This was addressed by Yuzbashyan, Baytin and Altshuler\cite{yba03}
who developed a strong coupling expansion based on the Bethe Ansatz solvability of the 
Richardson Hamiltonian. Thus the authors were able to compute the spectrum in orders of the
the inverse of the pairing interaction strength.

In this paper we complement the work of Ref. \onlinecite{yba03} by expanding around the same limit
but without relying on the Bethe Ansatz solvability of the Hamiltonian. As a result, perturbations
that destroy the Bethe Ansatz solvability of the model can also be considered. Our strategy is
to perform a so-called generalized Dyson boson-fermion mapping on the Hamiltonian. 
Generalized Dyson boson-fermion mappings,\cite{dsg93,ngd94,ngd95,ngd96,cg02} or Dyson mappings
for short, are instances of boson expansion methods.\cite{rs80,km91} A Dyson mapping
is an invertible linear transformation
that maps a many-fermion system onto a system in which independent bosons and fermions are present.
The bosons in the mapped system fulfill the role played by certain fermion pairs in the original
system. For the Richardson Hamiltonian, zero-momentum spin singlet Cooper pairs are thus represented
by bosons.

The paper has three main aims:
\begin{enumerate}
\item{We want to extend the work of Yuzbashyan {\it et al} and build further on the intuitive picture
for the low energy physics of superconducting nanograins by considering perturbations such as
an external electric field that breaks the Bethe Ansatz solvability. We consider not only
the spectrum, but also perform (zero-temperature) linear response calculations for time-dependent
perturbations.}
\item{Superconductivity is often thought of as the Bose-Einstein condensation of Cooper pairs.
However, Cooper pairs are not in themselves ``fundamental'' particles and can fairly easily be broken 
up into their constituent electrons by external perturbations. Viewing the superconducting state
as a condensate of non- or weakly interacting bosons leads one into the temptation to suppose
that the break-up of a Cooper pair is an event that does not depend on the state of the rest of
the system. We point out that this picture is wrong due to the exclusion principle. Because 
Cooper pairs consist of fermions, a high density of Cooper pairs reduces the number of unoccupied  
single particle orbitals into which the electrons resulting from Cooper pair break-up can decay.
As a result, the linear response of the system to a pair-breaking perturbation is not extensive
in the number of Cooper pairs.}
\item{We hope to convince readers of the utility and elegance of the Dyson Mapping as a tool for
analysing the Richardson Model. To this end we uncover a selection rule for the matrix elements of
single particle operators between eigenstates of the strong-pairing limit using the mapping. 
This selection rule seems to have gone unnoticed until now, posibly contributing to the assertion 
in Ref. \onlinecite{yba03}
that a conventional perturbation analysis (not exploiting Bethe Ansatz solvability) is not 
feasible. Also, we show how the mapping simplifies the task of calculating certain non-zero 
matrix elements of one-body operators between correlated many-body states.}
\end{enumerate} 

The paper is structured as follows. Section \ref{sect1} takes care of preliminaries.
The Richardson Hamiltonian and the Dyson mapping are introduced. The Hamiltonian obtained from
the Dyson mapping when transforming (or mapping) the Richardson Hamiltonian is presented. In Section
\ref{sect2} we motivate our claim that the Dyson mapping facilitates an analysis of the Richardson
Hamiltonian when the pairing term dominates the one-body term. In Section \ref{sect3}, a
perturbative diagonalization in orders of the kinetic energy is discussed. We calculate
the first three corrections to the ground state energy, and compare the results to those
previously obtained. \cite{yba03} We also
calculate the ground state expectation value of an arbitrary additive one-body operator to
first order in the kinetic energy. Brief mention is made of results obtained for excited states.
Section \ref{sect4} is devoted to the study of time-dependent phenomena. The main calculation
in this section yields the linear response of the expectation value of an arbitrary additive
one-body operator to an external perturbation. An appendix contains a detailed 
exposition of Rayleigh-Schr\"odinger perturbation theory for operators whose unperturbed spectra 
are degenerate. The formalism is general enough to hold for non-hermitian operators and
is presented in a way that makes it clear how to proceed in principle to arbitrarily high 
order corrections. 

Surprisingly, we could not locate a ready source in the existing literature that
presents degenerate perturbation theory in sufficient detail and general enough for present purposes. 
We have therefore included a more complete discussion in the
appendix than is strictly necessary for the derivations in the main body of the paper.

\section{\label{sect1}Hamiltonian and mapping}
With the single particle states, $\fvec{\mu\sigma}$, $\mu=-j,-j+1,\ldots,j-1,j$; $\sigma=\pm$,
we associate fermion creation and annihilation operators $a_{\mu\sigma}^+$ and $a_{\mu\sigma}$,
that obey the usual anti-commutation relations. The index $\sigma$ distinguishes between spin up
and spin down, while $\mu$ refers to momentum or some other quantum number, such that the state
$\fvec{\mu,\sigma}$ transforms to $\fvec{-\mu-\sigma}$ under time-reversal. The operators 
$a_{\mu\sigma}^+$ and $a_{\mu\sigma}$ act on fermion Fock space where we denote the
vacuum by $\fvec{0}$. 

The Richardson Hamiltonian is constructed in the fermion space as follows. First an operator
\begin{equation}
S^+=\frac{1}{\sqrt{\Omega}}\sum_{\nu=-j}^{j}a_{\nu+}^+ 
a_{-\nu-}^+,\label{ric1}
\end{equation}
is defined, with $\Omega=2j+1$ the number of single-particle orbitals. 
This choice normalizes $S^+\fvec{0}$ to unity. With the aid of the operator $S^+$ and its
conjugate $S$, the Richardson Hamiltonian can be stated as
\begin{equation}
H=\sum_{\nu=-j}^{j}\epsilon_\nu \left(a_{\nu+}^+ a_{\nu+}+a_{-\nu-}^+a_{-\nu-}\right)
-G S^+ S.\label{ric2}
\end{equation}
It is time-reversal invariant and contains a one-body kinetic term and a two-body interaction
or pairing term.  

The Hamiltonian describes electron-electron interactions in weakly disordered metallic nano-grains.
\cite{kaa00,vdr01} 
For simplicity, we have omitted the spin-exchange interaction, that should in principle be present
as well; we are primarily interested in superconduction correlations. Since the spin-exchange term 
commutes with all other terms in the Hamiltonian, it is straight-forward to re-introduce it at a
later stage if necessary. Its ommision here does not represent a serious loss of generality.

The following mathematical features of the model are important.

\subsection{The Seniority Limit}
If the kinetic term is switched off, the Richardson model reduces to Racah's
seniority model, so that we call this the seniority limit. The spectrum is 
explicitly known in the seniority limit.\cite{thou61,bro67} 
For a system with $2N$ particles, (with $N$ an integer),
the eigen-energies coalesce into $N+1$ levels
\begin{equation}
E^{(0)}_s=-\frac{G}{\Omega}(N-s)(\Omega+1-N-s),\label{ric6}
\end{equation}
with $s=0,1,\ldots,N$. Here the superscript is intended to indicate that this is the spectrum
in the absence of the kinetic term. The ground-state, with $s=0$ and energy
$E_0^{(0)}=-GN(\Omega+1-N)/\Omega$ is non-degenerate. All other levels are highly degenerate.
The energy difference (known as the superconducting gap) between the ground state and the 
(degenerate) $s=1$ level of eigenstates is $G$.
Up to normalization, the ground-state eigenvector is $\left(S^+\right)^N\fvec{0}$. 

\subsection{Bethe-ansatz Solvability}
It is possible to diagonalize the Richardson Hamiltonian beyond the seniority limit by
exploiting the fact that it is Bethe-ansatz solvable. See for instance Appendix B of Ref. 
\onlinecite{vdr01}. We focus on diagonalization in the subspace 
of time-reversal invariant states. These are states for which no single particle state $(\mu\sigma)$
is occupied without the time-reversed state $(-\mu-\sigma)$ also being occupied. It can easily
be shown that the Richardson Hamiltonian leaves this space invariant, and also that, if we know
how to diagonalize the Richardson Hamiltonian in the time-reversal invariant subspace, the 
diagonalization
in the whole fermion space can be achieved without further complication, thanks to the so-called
blocking effect. 

As a result of its Bethe-anzatz solvability, the following holds for the
eigenstates and eigen-energies of the Richardson Hamiltonian in the time-reversal invariant 
subspace: for a system of $2N$ fermions, let $\left\{E_k\right\}_{k=1}^N$ be a set of complex numbers that
satisfy the so-called Richardson equations
\begin{equation}
-\frac{\Omega}{G}=\sum_{\nu=-j}^j\frac{1}{E_k-2\epsilon_\nu}
+\sum_{l=1\not=k}^N\frac{1}{E_l-E_k},\label{ric3}
\end{equation}
for $k=1,2,\ldots,N$.
It can be shown that there are $\Omega\choose N$ such sets $\left\{E_k\right\}_{k=1}^N$, 
one for each dimension of the time-reversal invariant subspace. To
each solution corresponds an eigenvector of the Richardson Hamiltonian, given by 
\begin{equation}
\left|E\right>=\prod_{k=1}^N\left(\sum_{\nu=-j}^j\frac{a_{\nu+}^+ a^+_{-\nu-}}
{2\epsilon_\nu-E_k}\right)\left|0\right>,\label{ric4}
\end{equation}
up to normalization. The energy associated with this eigenstate is
\begin{equation}
E=\sum_{k=1}^N E_k.\label{ric5}
\end{equation}
Actually finding all the sets $\left\{E_k\right\}_{k=1}^N$ that solve the Richardson equations 
amounts to finding the simultaneous roots of $N$ polynomials of $N$ variables and degree 
$\Omega+N$. If either the number of single particle orbitals ($\Omega$) or the number of fermions
($2N$) becomes too large, or if analytical rather than numerical results are required, 
the Richardson equations become an obstacle.

\subsection{Strong Coupling Expansion}
As a consequence, few explicit results have been derived for this model, and most studies involve
numerical results based on the exact solution. To redress this state of affairs, Yusbashyan,
Baitin and Altshuler (YBA)recently developed a perturbation expansion in inverse powers of the
coupling constant $G$. With this expansion analytical results for the spectrum can be obtained
away from the thermodynamic limit, $N,\;V\rightarrow \infty$, where the discreteness
of one particle levels becomes important. In principle, the strong pairing limit, considered
by these authors, is realized in a small grain of ideal regular shape so that single electron
levels are highly degenerate. In the case where the grain is so small that the energy distance
between degenerate many-body levels is much larger than $G$, only the partially filled Fermi
level is relevant. The kinetic term is simply a constant proportional to the total number of
particles, and can be set equal to zero. Deviations from a perfectly regular shape slightly
lifts the degeneracy of the Fermi level and is incorporated by treating the kinetic term as
a perturbation.

Of course, in an irregularly shaped grain, the kinetic term cannot be viewed as a small perturbation
to the pairing term. However, even in this regime a perturbative analysis of the strong
coupling limit provides an intuitive insight into the nature of superconducting correlations.
Whereas the pairing term cannot be treated perturbatively, because the radius
of convergence for the expansion shrinks to zero in the thermodynamic limit, the $1/G$ expansion has a finite
convergence radius. Evidence suggests that for realistic $1/G$ 
the system behaves qualitatively similar to the small $1/G$ limit.

In the next sections, we will do such a $1/G$ perturbative analysis of the model. Our expansion method
complements that of Ref. \onlinecite{yba03}, in that it does not rely on the Bethe-Ansatz diagonalization of the Hamiltonian.
Thus, for instance, our method remains applicable when the kinetic term is replaced by an arbitrary
one-body perturbation, even though the resulting Hamiltonian can no longer be diagonalized using the
Bethe-Ansatz.

\subsection{Mapping the Building Blocks of the Model}
Our startegy involves mapping the Richardson Model onto
an equivalent boson-fermion model, by means of the so-called Dyson mapping. 
The mapping uncovers some hidden features
of the model and simplifies the task of calculating matrix elements of observables between correlated many-body
states. (See the Refs. \onlinecite{cg02, mythesis} for a more detailed discussion of the properties of the 
Dyson mapping.)

The operators $S$ and $S^+$ are pseudo-spin ladder operators for an $SU(2)$ irreducible
representation of dimension $\Omega$, or in other words of total pseudo-spin $j$.
\cite{yba03,thou61,bro67,and58} As such, they
have a lot in common with the boson creation and annihilation operators $B^\dagger$ and $B$,
especially if $\Omega$ is large. The Dyson mapping is a linear invertible operator 
designed to replace the creation (annihilation)
of Cooper pairs with the opperator $S^+$ ($S$), with the creation (annihilation) of bosons. 
Without showing how the mapping is derived we simply 
give the (linear extensions of) the Dyson images of the components needed to construct a system
described by the Richardson Hamiltonian. (See Ref. \onlinecite{mythesis} for a complete discussion 
on how the Dyson mapping is defined as well as a derivation of the images of the operators that
constitute the building blocks of the Richardson Hamiltonian.)
\begin{widetext}
\begin{eqnarray}
\left(S\right)_D&=&B,\label{map6}\\
\left(S^+\right)_D&=&B^\dagger\left(1-\frac{\nf+\nb}{\Omega}\right),\label{map7}\\
\left(a_{\nu\sigma}\right)_D&=&\alpha_{\nu\sigma}+\sigma\sqrt{\Omega}B\alpha_{-\nu-\sigma}^\dagger
\frac{1}{\Omega-\nf}\nonumber\\
&&\hspace{20mm}+\mcs^\dagger B \alpha_{\nu\sigma}\frac{\Omega}{(\Omega-\nf)(\Omega-\nf+1)},
\label{map8}\\
\left(a_{\nu\sigma}^+\right)_D&=&\alpha_{\nu\sigma}^\dagger\frac{\Omega-\nf-\nb}{\Omega-\nf}
+\frac{\sigma}{\sqrt{\Omega}}B^\dagger\alpha_{-\nu-\sigma}\nonumber\\
&&\hspace{20mm}-\sigma\sqrt{\Omega}\mcs^\dagger \alpha_{-\nu-\sigma}
\frac{\Omega-\nf-\nb}{(\Omega-\nf)(\Omega-\nf+1)}.\label{map9}
\end{eqnarray}
\end{widetext}
In these expressions, $\alpha^\dagger_{\nu\sigma}$ and its conjugate are fermion operators, with
the usual anti-commutation relations. We will refer to the the particles created and annihilated
by them as ideal fermions, to distinguish them from the fermions of the unmapped system. $B^\dagger$
and $B$ are boson operators with the usual commutation relations. Importantly, the boson operators
commute with the ideal fermion operators.
Furthermore, $\nf$ counts the number of ideal fermions, $\nb$ counts the number of bosons
and $\mcs^\dagger$ is defined as the operator $S^+$ with the real fermion operators 
$a_{\mu\sigma}^+$ replaced by ideal fermion operators $\alpha_{\mu\sigma}^\dagger$. The fermion
vacuum $\fvec{0}$ maps onto the boson-fermion vacuum $\bfvec{0}$, while the functional $\ffun{0}$, 
maps onto a functional $\bffun{0}$. 

Note two subtleties of the mapping. Firstly it is not unitary, in the sense that 
$\left(X\right)_D^\dagger\not=\left(X^+\right)_D$. (We denote hermitian conjugation in the 
fermion space with a $+$ sign while in the boson fermion space we denote it with a $\dagger$,
to emphasize that the mapping of the conjugate is not the same as the conjugate of the mapping.)
This implies that a hermitian fermion operator
such as $H$ will in general be mapped onto a non-hermitian operator $\left(H\right)_D$. The equivalence
between the Richardson Hamiltonian and the boson-fermion system is therefore a very formal mathematical
one. It is not the physical equivalence between two systems that can both be realized in a (gedanken)
laboratory. Secondly, the reader will have noticed the word ``linear extension'' used in connection
with the Dyson images of fermion operators. The need for such terminology can be appreciated as follows:
Recall first that the fermion Hilbert space on which the Richardson Hamiltonian is defined, was built up
from a $2\Omega = 4j+2$ dimensional one-particle Hilbert space. The operators used in the construction of the 
Richardson Hamiltonian therefore act on a finite dimensional vector space. Yet the Dyson image of 
the fermion operator $S$ for instance is the boson-operator $B$. While $S$ acts in a finite vector space,
$B$ acts in an infinite-dimensional vector space: an infinite number of bosons can occupy the same state.
What is happening here is that the operators in (\ref{map9}) leave a subspace of the full infinite
dimensional boson-fermion space invariant. Within this subspace the operators of (\ref{map9}) are equivalent
to the building blocks of the Richardson Hamiltonian. One might naively suppose that the physical subspace 
of boson fermion space is that in which twice the number of bosons plus the number of ideal fermions equal
the number of real fermions. However, counting dimensions one sees that the dimension of
this subspace of boson fermion space is still too large. The physical subspace is still smaller and its
identification more subtle, but fully solved.\cite{mythesis} 
While this brings about the complication that one
should take care to choose initial and final states from this subspace, working in a larger than necessary
Hilbert-space has advantages as well, as will become clear in the next section.  

At this point it is still not obvious what one gains by performing a Dyson mapping on the Richardson Hamiltonian.
This is addressed more fully in the next section. Note here, however, that the mapped operators have
the following non-trivial but favorable properties:
\begin{enumerate}
\item{A state containing $2N$ real fermions maps onto a state for which 
twice the number of bosons plus the number of ideal fermions is equal to $2N$.} 
\item{The operator $\left(a_{\mu\sigma}\right)_D$ only contains terms that annihilate an ideal fermion
without changing the number of bosons or that annihilate a boson and create an ideal fermion. 
In fact, with the annihilation of a real fermion is associated three coherent processes in the mapped system.
In the first an ideal fermion is annihilated, in the second a boson is annihilated while an ideal fermion is
created, and in the third an ideal fermion and a boson is annihilated while an ideal fermion pair is 
created.
No terms that change the boson number by more than one occur. Similarly, the operator 
$\left(a_{\mu\sigma}^+\right)_D$ can create an ideal fermion without changing the number of bosons
or it can annihilate an ideal fermion and create a boson, but contains no terms that change the
boson number by more than one.} 
\item{Also, a real fermion many-body state in which all fermions occur in time-reversed pairs
$a^+_{\mu+}a^+_{-\mu-}$ maps onto a boson-fermion state in which ideal 
fermions only occur in time-reversed pairs $\alpha^\dagger_{\nu+}\alpha^\dagger_{-\nu-}$.} 
\item{A real
fermion state for which the number of real spin up fermions minus the number of real 
spin down fermions is $M$, maps onto a boson-fermion state for which the number of ideal spin
up fermions minus the number of ideal spin down fermions is also $M$.} 
\end{enumerate}
We stress that the extensions of the
images of the operators that appear above are exact and complete as they stand. The expressions are not
the first few terms in series of which `higher order' terms have been neglected, nor was any other approximation
made. 

These building blocks can be combined to find the linear extension $H_D$ of the image of the 
Richardson Hamiltonian. For the pairing term we simply get
\begin{equation}
P_D=-G\left(S^+S\right)_D=-\frac{G}{\Omega}\nb(\Omega+1-\nb-\nf).\label{map9b}
\end{equation}
Instead of immediately focusing on the image of the kinetic term in
the Richardson Hamiltonian, we rather consider an arbitrary additive one-body operator 
\begin{equation}
A=\sum_{\mu\nu=-j}^j A_{\mu\nu}\left(a_{\mu+}^+a_{\nu+}+a_{\mu-}^+a_{\nu-}\right),
\label{map9c}
\end{equation}
that leaves the fermion space invariant \footnote{This becomes an issue when the finite set of 
one-particle indices $\mu=-j,\ldots,j$ is a subset of a larger (possibly infinite) set of indices. 
See for instance the example of the particles confined to a ring, discussed in Section \ref{sect4}.}
and does not distinguish between spin up and spin down 
states. These two restrictions are not fundamental; they are imposed purely for the sake of simplicity.
The linear extension of the image of $A$ under the Dyson mapping can be conveniently
separated into three parts $A_D=A_{D-}+A_{D0}+A_{D+}$, for which we have the following explicit
expressions.
\begin{widetext}
\begin{eqnarray}
A_{D-}&=&B^\dagger\xi_A\label{map10}\\
A_{D0}&=&\mcq-\frac{\nb}{\Omega-\nf}(\mcq+\mcq_*-2\tr\mcq)-\mcs^\dagger\xi_A
\frac{(\Omega-2\nb-\nf)\Omega}{(\Omega-\nf)(\Omega-\nf+2)}\label{map11}\\
A_{D+}&=&\xi_A^\dagger B\frac{(\Omega-\nf-\nb)\Omega}{(\Omega-\nf-1)(\Omega-\nf)}
+\mcs^\dagger B(\mcq+\mcq_*-2\tr\mcq)\frac{(\Omega-\nb-\nf)\Omega}
{(\Omega-\nf-1)(\Omega-\nf)^2}\nonumber\\
&&\hspace{2mm}-(\mcs^\dagger)^2B\xi_A\frac{(\Omega-\nb-\nf)\Omega^2}{(\Omega-\nf-1)
(\Omega-\nf)^2(\Omega-\nf+1)}\label{map12}
\end{eqnarray}
\end{widetext}
In these expressions, the operator $\mcq$ is defined, similar to $\mcs^\dagger$, as the 
operator $A$, with the real fermion operators $a_{\mu\sigma}^+$ and $a_{\mu\sigma}$ replaced
by ideal fermion operators $\alpha_{\mu\sigma}^\dagger$ and $\alpha_{\mu\sigma}$. The operator
$\mcq_*=\sum_{\mu\nu\sigma} A_{-\nu-\mu}\alpha_{\mu\sigma}^\dagger
\alpha_{\nu\sigma}$ is the operator into which $\mcq$ transforms under time-reversal. The operator
$\xi_A^\dagger$ is given by
\begin{eqnarray}
\xi_A^\dagger&=&\comm{\mcq,\mcs^\dagger}\nonumber\\
&=&\frac{1}{\sqrt{\Omega}}\sum_{\mu\nu=-j}^j\left(A_{\mu\nu}+A_{-\nu-\mu}\right)
\alpha_{\mu\nu}^\dagger\alpha_{-\nu-\mu}^\dagger,\label{map15}
\end{eqnarray}
and $\xi_A$ is its hermitian conjugate in the mapped boson-fermion space $\bfspace$. With $\tr\mcq$ we mean the trace
$\sum_{\mu=-j}^j A_{\mu\mu}$ of the one-body matrix $A_{\mu\nu}$. The separation of $A_D$ into
three terms has been done in a way such that $A_{D-}$ annihilates two ideal fermions and
creates a boson while  $A_{D0}$ leaves both the number of bosons and ideal fermions unchanged 
and $A_{D+}$ annihilates a boson and creates two ideal fermions. 

From (\ref{map10})-(\ref{map12}) it is clear that an additive one-body operator 
$A=\sum_{\mu\nu\sigma}A_{\mu\nu}a_{\mu\sigma}^+ a_{\nu\sigma}$ that is odd under 
time-reversal, i.e. $A_{\mu\nu}=-A_{-\nu-\mu}$, is simply mapped onto an operator 
$A_D=\mcq=\sum_{\mu\nu\sigma}A_{\mu\nu}\alpha_{\mu\sigma}^\dagger \alpha_{\nu\sigma}$. (Remember
that hermitian operators that are odd under time-reversal are traceless.) This has the
consequence that $A_D$ commutes with $P_D=-G\left(S^+S\right)_D$. 
That additive one-body operators that are odd under time-reversal commute with the pairing 
interaction, can also be checked without difficulty in the original system.

To summarize the most important points of this section then, 
there is a similarity transformation $\dmap$, known as the Dyson mapping, 
that maps many-fermion states  onto states in which Cooper pairs are
replaced by bosons. Any fermion operator $X$ transforms into
an operator $\dmap X \dmap^{-1}$, defined on a subspace of the full boson-fermion
Hilbert space  and is
said to be the image of $X$. We have presented linear extensions of the images of the
building blocks of the Richardson Hamiltonian to the whole
boson fermion space. We denote the extended image of $X$ under the Dyson mapping as 
$X_D$. Somewhat carelessly we often call $X_D$ the
image of $X$ under the Dyson mapping. Bra's and kets in the original fermion Hilbert space are written
with angular brackets, while bra's and kets in the new boson-fermion Hilbert space  are written
with round brackets.

From the images of the operators in (\ref{map6})-(\ref{map9}) the
the image of the Richardson Hamiltonian under the Dyson mapping can be constructed (\ref{map9b}),
(\ref{map10})-(\ref{map12}). In the next section we explain why it is useful to transform
the Richardson Hamiltonian using the Dyson mapping if the pairing term dominates the one-body
term.

\section{\label{sect2}The Utility of the Dyson Mapping}
Let us diagonalize the image $P_D=-G\left(S^+S\right)_D$ of the pairing term, supposing that
the original system contains $2N$ fermions. This implies  
that twice the number of bosons plus the number of ideal fermions equal $2N$ in the mapped system. 
Since the mapped pairing term (\ref{map9b}) is clearly diagonal in boson-fermion space, 
we immediately find by simple inspection
that any state in boson-fermion space that contains $N-s$ bosons and $2s$ ideal fermions is an eigenstate
of $P_D$, with eigenvalue $E_s^{(0)}=-G(N-s)(\Omega+1-N-s)/\Omega$. This is exactly the
same spectrum as that of the seniority model $P=-GS^+S$ on the fermion side, as is of course to be expected from
a faithful mapping. 
However, if one counts the degeneracy of many-body levels, one finds that the
mapped pairing term $P_D$ has a higher degeneracy than the original pairing term $P$. For instance
the degeneracy of the first level of excitations ($s=1$) for $P_D$ is $2\Omega^2-\Omega$ while for
$P$ it is $2\Omega^2-\Omega-1$. This discrepancy is due to the fact that $P_D$ is defined
in a larger Hilbert space than $P$. One of the eigenstates in the $s=1$ level of the spectrum
of $P_D$ is unphysical. The identification of this unphysical state and of all unphysical states
in the other levels of the spectrum of $P_D$ has been carried out elsewhere [\onlinecite{mythesis}, p.\ 99]. The subspace of
boson-fermion space on which the Dyson mapping establishes an equivalence with the Richardson Hamiltonian
is completely known. For the purposes of the present text a full knowledge of this physical subspace is however not necessary. 

\subsection{An Important Selection Rule}
We are now ready to state one of the main new technical results of our investigation which subsequently
also turns out to be of great utility.
Every eigenstate (and therefore every physical eigenstate)
of $P_D$ with energy $E^{(0)}_s$ contains $N-s$ bosons and 
$2s$ ideal fermions. 
As a remarkable 
consequence, the boson-fermion counterpart $A_D$ of a one-body operator $A$ only has non-zero matrix
elements connecting states in a level $E^{(0)}_s$ with states in the levels $E^{(0)}_{s'}$ with $s'=s-1,s,s+1$. This
is because $A_D$ does not change the number of ideal fermions by more than two. Because of the
linearity and invertibility of the Dyson mapping, this result is
equally true in the original system: 
\begin{equation}
\left<E_s^{(0)}\right|X\left|E_{s'}^{(0)}\right>=0
\end{equation}
unless $s'\in{s-1;s;s+1}$, where $\left|E_{s}\right>$ is any eigenstate from level $s$ of the pairing operator
$P=-GS^+S$ and similarly $\left|E_{s'}^{(0)}\right>$ comes from level $s'$. $X$ is any additive one-body 
fermion operator. 
Since this useful selection rule is not as immediately apparent in the original system, it seems 
to have gone unnoticed in the 
literature. With its aid the number of matrix-elements in a perturbation expansion for which
non-trivial calculations are required is reduced by orders of magnitude.
\subsection{The Advantage of a Larger Hilbert Space}
When calculating matrix elements of operators, one often needs a spectral representation of the
identity operator, or of a projection operator. (If $\left\{\left|k\right>\right\}_{k=1}^M$ is an orthogonal
basis for a vector space then $I=\sum_{k=1}^N\left|k\right>\left<k\right|$ is a spectral representation
of the identity operator.) 

Now consider the somewhat artificial but helpful example of a three-dimensional
vector space. Imagine that we are working in a 2-dimensional sub-space spanned by the two orthogonal vectors
$\left|1\right>=\sin\theta\;\sin\phi\left|x\right>+\sin\theta\;\cos\phi\left|y\right>
+\cos\theta\left|z\right>$ and $\left|2\right>=\cos\phi\left|x\right>-\sin\phi\left|y\right>$.
For vectors $\left|\alpha\right>$ and $\left|\beta\right>$ in this subspace, the inner product 
$\left<\alpha\right.\left|\beta\right>$ can be written as $\left<\alpha\right|I\left|\beta\right>$.
The identity operator $I$ has the decomposition
\begin{eqnarray}
I&=&\left|1\right>\left<1\right|+\left|2\right>\left<2\right|\nonumber\\
&=&(\sin^2\theta\;\sin^2\phi+\cos^2\phi)\left|x\right>\left<x\right|\nonumber\\
&&-\cos^2\theta\sin\phi\cos\phi\left|x\right>\left<y\right|+\mbox{ seven similar terms}.
\end{eqnarray}
Instead of this complicated expression, we are free to use the identity operator defined for the whole
three dimensional space $I_{\rm extended}=\left|x\right>\left<x\right|+\left|y\right>\left<y\right|
+\left|z\right>\left<z\right|$, since $\left<\alpha\right|I\left|\beta\right>=\left<\alpha\right|I_{\rm extended}
\left|\beta\right>$ for all $\left|\alpha\right>$ and $\left|\beta\right>$ in the two-dimensional subspace.

This same advantage is to be had by working with the mapped instead of the original system:
In any perturbation expansion where the pairing term is perturbed by a one-body operator $A$,
typically quantities like 
$\ffun{\psi' s'}A\Pi_{s_1}A\Pi_{s_2}\ldots\Pi_{s_n}A\fvec{\psi s}$ must be calculated.
Here $\Pi_s$ is the projection
operator that singles out the component of the state it acts on in the space spanned by 
vectors that lie in level $s$ of the spectrum of the unperturbed Hamiltonian. The vectors 
$\fvec{\psi s}$ and $\fvec{\psi' s'}$ denote eigenstates
from the levels $s$ and $s'$ respectively. 

For simplicity, let us focus on the modestly non-trivial
quantity $\ffun{GS}A\Pi_{1}A\fvec{GS}$, which is encountered in the calculation of the second
order correction to the ground state in the strong pairing limit. 
Here $\fvec{GS}\propto \left(S^+\right)^N\fvec{0}$ is the normalized ground state in the
seniority limit. To make any head-way, an expression must be found for $\Pi_1$. To do this, we
need a complete basis for the $s=1$ eigenspace. Finding such a basis is already a
non-trivial task, and becomes progressively harder for larger values of $s$. It can be shown
that any complete
basis for the $s=1$ eigenspace can be written in the form 
$\left\{\fvec{k}=\left(S^+\right)^{N-1}\sum_{\mu\nu=-j}^j c_{k\mu\nu}
a_{\mu+}^+a_{\nu-}^+\left|0\right>\right\}_{k=1}^{\omega_1}$, where $\omega_1$ is the degeneracy
of the level $s=1$ and $c_{k\mu\nu}$ are complex numbers. 
(We only consider states with equal numbers of spin up and spin down fermions
because the ground state contains as many fermions with spin up as spin down, and we have 
conveniently assumed $A$ not to introduce spin flipping.) The projection operator $\Pi_1$ can
then be written as $\Pi_1=\sum_{k=1}^{\omega_1}\fvec{k}\ffun{k}$. This leaves us with the 
unpleasant task of calculating matrix elements such as
\begin{widetext}
\begin{equation}
\ffun{GS}A\fvec{k}\nonumber\\
=\ffun{0}S^N\left(\sum_{\mu\nu\sigma}A_{\mu\nu}a_{\mu\sigma}^+a_{\nu\sigma}
\right)\left(S^+\right)^{N-1}\left(\sum_{\rho\lambda}c_{k\rho\lambda}a_{\rho+}^+a_{\lambda-}^+
\right)\fvec{0}.
\label{util1}
\end{equation}
\end{widetext}
In order to do this calculation, 
the operators must be normal-ordered. 
As far as we can see, the operators $A$, $S$, $S^+$ and
$\sum_{\mu\nu}c_{k\mu\nu}a_{\mu+}^+a_{\nu-}^+$ do not belong to any useful algebra that would 
simplify the normal ordering.
This means that, in the myriad commutation operations
involved in normal ordering, new linear combinations of fermion operators
that have not been encountered before are frequently
generated. The result is that the number of non-trivial operations involved in calculating
(\ref{util1}) grows explosively as $N$ increases. 

Let us now consider the calculation of $\ffun{GS}A\Pi_1A\fvec{GS}$, using the Dyson mapping.
Firstly, by replacing $S^+$ with $\left(S^+\right)_D$ and $S$ with $S_D$ we find that 
\begin{equation}
\dmap \fvec{GS}=\mathcal C \frac{\left(B^\dagger\right)^N}{\sqrt{N!}}\bfvec{0}\equiv\mathcal C
\bfvec{-},\label{util1a}
\end{equation}
 and
\begin{equation}
\ffun{GS}\dmap^{-1}=\tilde{\mathcal C} \bffun{0}\frac{B^N}{\sqrt{N!}}\equiv\tilde{\mathcal C}
\bffun{-},\label{util1b}
\end{equation} 
where $\mathcal C$ and $\tilde{\mathcal C}$ are complex numbers. The notation is intended to
indicate that $\bfvec{-}$ contains no ideal fermions. Because it holds that
$1=\fxpv{GS|GS}=\ffun{GS}\dmap^{-1}\dmap\fvec{GS}$, we know that $\tilde{\mathcal C}\mathcal C=1$.
Thus we find 
\begin{equation}
\ffun{GS}A\Pi_1 A\fvec{GS}=\bffun{-}A_{D-}A_{D+}\bfvec{-}.\label{util2}
\end{equation}
In this expression, the image of $\Pi_1 A\fvec{GS}$ is $\mathcal C A_{D+}\bfvec{-}$. The only
part of $\ffun{GS}A\dmap^{-1}$ that has a non-zero overlap with this is $\tilde{\mathcal C}
\bffun{GS}A_{D-}$. To compute  $\bffun{-}A_{D-}A_{D+}\bfvec{-}$, it is 
convenient to insert the
identity operator in the $s=1$ eigenspace (which is just a linear extension of $\dmap \Pi_1\dmap^{-1}$)
between $A_{D-}$ and $A_{D+}$. If we set $\bfvec{\mu\nu_*}=
\frac{1}{\sqrt{(N-1)!}}\left(B^\dagger\right)^{N-1}
\alpha_{\mu+}^\dagger\alpha_{-\nu-}^\dagger\bfvec{0}$ then
\begin{equation}
\bffun{-}A_{D-}A_{D+}\bfvec{-}=\sum_{\mu\nu=-j}^j\bffun{-}A_{D-}\bfvec{\mu\nu_*}\bffun{\mu\nu_*}
A_{D+}\bfvec{-}.\label{util3}
\end{equation}
Here the notation $\bfvec{\mu\nu_*}$ is intended to indicate that the state contains two ideal
fermions (and therefore $N-1$ bosons), and further that one of the ideal fermions occupies the state
labeled by $(\mu,+)$ for which $\mu$ is shorthand, while the other occupies the state $(-\nu,-)$
for which $\nu_*$ is shorthand.
The matrix-elements in the above expression present no great challenge to compute. This is 
due to the fact that the bosons and ideal fermions have simpler commutation rules than the
building blocks of states and operators on the fermion side. Particularly, ideal fermion operators
commute with boson operators, as opposed to real fermion operators that have non-trivial
commutators with the $S^+$ and $S$ operators. It is true that things become 
significantly more complicated in the mapped system for
levels $s$ higher up in the spectrum. However, there is no ceiling or clear boundary beyond which
calculations suddenly become impossibly complicated. For simplicity we will only calculate 
quantities that involve matrix-elements between states from the levels $s=0$ and $s=1$ in this
paper, but calculations involving states from the level $s=2$ are no less feasible. Only, the
calculations become tedious and the results too intricate for casual interpretation. 

We now write down all the matrix-elements that we are going to use in the rest of this paper.
\begin{widetext}
\begin{eqnarray}
\bffun{-}A_{D0}\bfvec{-}&=&2N\frac{\tr A}{\Omega},\label{util7a}\\
\bffun{-}A_{D-}\bfvec{\mu\nu_*}&=&\sqrt{\frac{N}{\Omega}}\left(A_{\nu\mu}+A_{-\mu-\nu}\right),
\label{util7b}\\
\bffun{\mu\nu_*}A_{D+}\bfvec{-}&=&\sqrt{\frac{N}{\Omega}}\frac{\Omega-N}{\Omega-1}
\left(A_{\mu\nu}+A_{-\nu-\mu}-2\delta_{\mu\nu}\frac{\tr A}{\Omega}\right),\label{util7c}\\
\bffun{\mu'\nu_*'}A_{D0}\bfvec{\mu\nu_*}&=&
2\frac{N-1}{\Omega-2}\tr A\,\delta_{\mu'\mu}\delta_{\nu'\nu}
+\left\{A_{-\nu'-\nu}-\frac{N-1}{\Omega-2}\left(A_{-\nu'-\nu}
+A_{\nu\nu'}\right)\right\}\delta_{\mu'\mu}\nonumber\\
&&\hspace{20mm}+\left\{A_{\mu'\mu}-\frac{N-1}{\Omega-2}\left(A_{\mu'\mu}
+A_{-\mu-\mu'}\right)\right\}\delta_{\nu'\nu}\nonumber\\
&&\hspace{50mm}-\frac{\Omega-2N}{\Omega(\Omega-2)}
\left(A_{\nu\mu}+A_{-\mu-\nu}\right).\label{util7}
\end{eqnarray}
\end{widetext}
Calculating the corresponding many-body matrix-elements in the unmapped system seems intractable to us.

\section{\label{sect3}Time-independent Perturbation Theory}
In this section we discuss the implementation of Rayleigh-Schr\"odinger (time-independent)
perturbation theory for the linear extension $H_D$ of the image of the Richardson Hamiltonian.
The image $K_D$ of the kinetic energy operator is treated as a perturbation to the 
image $P_D$ of the pairing operator.
The original kinetic energy operator $K$ defined on fermion space is an additive 
one-body operator with one-body matrix elements $K_{\mu\nu}=\delta_{\mu\nu}\epsilon_\mu$.
It is assumed to be time-reversal invariant, i.e. $\epsilon_{\mu}=\epsilon_{-\mu}$.
We focus on the first few corrections to the ground state energy and eigenvector. Brief mention
is made of the procedure involved in the expansion for the degenerate excited states.

In a recent paper by Yuzbashyan and co-workers,\cite{yba03} a procedure was
presented for determining the eigenvalues perturbatively. The strategy
there was to develop an expansion for $\sum_{k=1}^N E_k$ where $E_k$, $k=1,\ldots,N$ satisfy the
Richardson equations (\ref{ric3}). 
We do not claim that the method described in this section is easier
to implement than the method of Ref.\onlinecite{yba03}
if one only wants to calculate corrections to the energies. What we do claim is that
Rayleigh-Schr\"odinger perturbation theory is simpler to implement
than the authors of Ref.\onlinecite{yba03} may have realized. This we attribute to the structure resulting
from the Dyson mapping., as we proceed to show.
With our procedure, it is feasible 
to calculate the first few corrections to eigenstates at the bottom of the spectrum. Thus, our
procedure allows one to compute, for instance, the first few corrections to the ground state 
expectation values of observables.

\subsection{Ground State Energy}
As a demonstration that the formulas of the previous 
section are valid, we will start by calculating the first three corrections to the ground-state
energy, and compare the results with those obtained by Yuzbashyan and co-workers. The general 
formulas for 
these corrections (from the appendix) are
\begin{eqnarray}
\Delta_{GS}^{(1)}&=&\bffun{-}K_D\bfvec{-},\label{tip1}\\
\Delta_{GS}^{(2)}&=&\bffun{-}K_DQ_{GS}K_D\bfvec{-},\label{tip2}\\
\Delta_{GS}^{(3)}&=&\bffun{-}K_DQ_{GS}\left(K_D-\Delta_{GS}^{(1)}\right)Q_{GS}K_D\bfvec{-}.
\label{tip3}
\end{eqnarray}
Here $Q_{GS}$ is a linear operator defined through its image on eigenstates of the mapped pairing operator 
$P_D$ as follows. 
\begin{equation}
Q_{GS}\bfvec{\psi,s}=\left\{\begin{array}{lcl}0&{\rm if}&s=0\\
\frac{1}{E_0^{(0)}-E_{s}^{(0)}}\bfvec{\psi,s}&{\rm if}&s>0\end{array}\right.\label{tip4},
\end{equation}
for any state $\bfvec{\psi,s}$ with $N-s$ bosons and $2s$ ideal fermions. Using the 
decomposition $K_D=K_{D-}+K_{D0}+K_{D+}$ together with this definition of $Q_{GS}$, 
we can replace (\ref{tip1})-(\ref{tip3}) by 
\begin{eqnarray}
\Delta_{GS}^{(1)}&=&\bffun{-}K_{D0}\bfvec{-},\label{tip5}\\
\Delta_{GS}^{(2)}&=&\bffun{-}K_{D-}K_{D+}\bfvec{-},\label{tip6}\\
\Delta_{GS}^{(3)}&=&\bffun{-}K_{D-}\left(K_{D0}-\Delta_{GS}^{(1)}\right)K_{D+}\bfvec{-}.
\label{tip7}
\end{eqnarray}
(Remember that $E_1^{(0)}-E_0^{(0)}=1$.)
Immediately, we have from (\ref{util7a}) that 
\begin{equation}
\Delta_{GS}^{(1)}=2N\overline{\epsilon},\label{tip8}
\end{equation}
where, in general,
$\overline{\epsilon^k}=\sum_{\mu=-j}^j\epsilon_\mu^k/\Omega$. In order to calculate 
$\Delta_{GS}^{(2)}$ and $\Delta_{GS}^{(3)}$, we insert appropriate representations of the identity
operator to reduce the expressions in (\ref{tip6}) and (\ref{tip7}) to sums over the matrix-elements in
(\ref{util7a})-(\ref{util7}). For the second order correction we find
\begin{equation}
\Delta_{GS}^{(2)}=\sum_{\mu=-j}^j\bffun{-}K_{D-}\bfvec{\mu\mu_*}\bffun{\mu\mu_*}K_{D+}
\bfvec{-},\label{tip9}
\end{equation}
where the matrix elements are $\bffun{-}K_{D-}\bfvec{\mu\mu_*}=2\sqrt{N/\Omega}
\epsilon_\mu$ and $\bffun{\mu\mu_*}K_{D+}\bfvec{-}=2\sqrt{N/\Omega}
\left(\epsilon_\mu-\bar{\epsilon}\right)$. The fact that, in the sum over states, only states $\bfvec{\mu\mu_*}$ 
whose ideal fermions form time-reversal invariant pairs appear, can be 
traced back to the time-reversal invariance of the original kinetic energy operator $K$. 
Hence we find
\begin{equation}
\Delta_{GS}^{(2)}=-\frac{4N(\Omega-N)}{G(\Omega-1)}\overline{\left(\epsilon-
\overline{\epsilon}\right)^2}.
\label{tip10}
\end{equation}
To calculate the third order correction, we need the additional matrix elements
$\bffun{\mu\mu_*}K_{D0}-\Delta_{GS}^{(1)}\bfvec{\nu\nu_*}=
2\frac{\Omega-2N}{\Omega-2}\left\{\left(\epsilon_\mu-\overline{\epsilon}\right)\delta_{\mu\nu}
-\frac{\epsilon_\nu}{\Omega}\right\}$. Thus we find
\begin{widetext}
\begin{eqnarray}
\Delta_{GS}^{(3)}&=&\frac{1}{G^2}
\sum_{\mu\nu=-j}^j\bffun{-}K_{D-}\bfvec{\mu\mu_*}\bffun{\mu\mu_*}K_{D0}-\Delta_{GS}^{(1)}
\bfvec{\nu\nu_*}\bffun{\nu\nu_*}K_{D+}\bfvec{-}\nonumber\\
&=&\frac{8N(\Omega-N)(\Omega-2N)}{G^2\Omega(\Omega-1)(\Omega-2)}
\sum_{\mu\nu=-j}^j\epsilon_\mu\left\{\left(\epsilon_\mu-\overline{\epsilon}\right)\delta_{\mu\nu}
-\frac{\epsilon_\nu}{\Omega}\right\}\left(\epsilon_\nu-\overline{\epsilon}\right)\nonumber\\
&=&\frac{8N(\Omega-N)(\Omega-2N)}{G^2(\Omega-1)(\Omega-2)}
\overline{\left(\epsilon-\overline{\epsilon}\right)^3}.\label{tip11}
\end{eqnarray}
\end{widetext}
After converting the notation used in this paper to the notation of Ref. \onlinecite{yba03}, namely
$N \rightarrow M$, $\Omega \rightarrow N$, $\overline{\epsilon^k}\rightarrow2^{-k}s_k/N$
and $G\rightarrow N\lambda d$, we find our first and second order corrections agreeing with
those in Ref. \onlinecite{yba03}. The 
seemingly different third order corrections are the result of a small arithmetic or typing error in
Ref. \onlinecite{yba03}. It is not hard to check explicitly that the method of Ref. \onlinecite{yba03}, when 
implemented correctly, does indeed yield the same answer as we get in (\ref{tip11})

\subsection{Grounstate Eigenvector and Expectation Values}
Let us now calculate the first order correction
to the ground-state eigenvector. (To us it is not clear whether this can be calculated using the method of
Ref. \onlinecite{yba03}.) Since the perturbation is not hermitian, a distinction must be
made between the left and right perturbed eigenvectors. (Such a distinction was not necessary for
the unperturbed ground-state as the unperturbed operator $P_D$ happened to be hermitian.) The
right ground-state eigenvector of the perturbed operator $K_D+P_D$ is, to first order,
$\bfvec{GS_R}=\bfvec{-}+\bfvec{GS^{(1)}_R}$ whereas the left ground state eigenvector is
$\bfvec{GS_L}=\bfvec{-}+\bfvec{GS^{(1)}_L}$, also to first order. Here $\bfvec{GS_L}$ is 
defined to be the ground-state eigenvector of $(K_D+P_D)^\dagger$.

According to the appendix, the first order correction for the right-eigenvector is
\begin{eqnarray}
\bfvec{GS^{(1)}_R}&=&Q_{GS}K_D\bfvec{-}\nonumber\\
&=&-\frac{1}{G}K_{D+}\bfvec{-}\nonumber\\
&=&2\sqrt{\frac{N}{\Omega}}\frac{\Omega-N}{\Omega-1}\sum_{\mu=-j}^j\frac{\overline{\epsilon}
-\epsilon_\mu}{G}\bfvec{\mu\mu_*},\label{tip12}
\end{eqnarray}
where the second line follows from the selection rule for the matrix-elements of the images of 
additive one-body operators explained in Section \ref{sect2}. Similarly we have for the left 
eigenvector
\begin{eqnarray}
\bfvec{GS^{(1)}_L}&=&Q_{GS}K_D^\dagger\bfvec{-}\nonumber\\
&=&-\frac{1}{G}K_{D-}^\dagger\bfvec{-}\nonumber\\
&=&-2\sqrt{\frac{N}{\Omega}}\sum_{\mu=-j}^j\frac{\epsilon_\mu}{G}\bfvec{\mu\mu_*}.\label{tip13}
\end{eqnarray}
At this point we emphasise that the eigenvectors found above are physical. If one were for instance to find the ground state perturbatively to first order for the original fermion Hamiltonian
and then apply the Dyson mapping, one would obtain the above right-eigenvector. In general,
implementation of exactly mapped operators obtained from the Dyson mapping do not {\it per se} introduce spurious components. It is only when the Hamiltonian is
diagonalized in the complete boson (or boson-fermion) Fock space, rather than in the physical subspace, that spurious states may be obtained. While the physical subspace may be obtained {\it
ab initio} via the mapping of a basis in the original space, this would nullify the relative simplicity obtained when using a standard basis for the complete boson (or boson-fermion) space.
Nevertheless, if the Hamiltonian is mapped exactly, as we do here,  (i.e. without any truncation of degrees of freedom), the resulting eigenstates neatly separate into completely physical states and states with 
spurious components, and a number of ways exist to distinguish these (see e.g.\ Refs.\ \onlinecite{geh86,dgh91}). For the Richardson Hamiltonian this is done in in a novel way -- see Ref.\ \onlinecite{mythesis}, p.\ 99. 
The neat division between completely physical states and states with spurious components still holds when 
the mapped Hamiltonian is diagonalized perturbatively in the kinetic term, as is explained in detail in the last paragraph of the appendix.

To calculate higher order corrections to these eigenvectors, we need to know matrix elements
between higher excited states. Specifically, for the second order corrections, matrix
elements of the kinetic term between states in the $s=1$ and $s=2$ levels are required. 
We will however not proceed beyond the corrections already calculated. 

Let us calculate the ground state expectation value for a one-body additive operator $A$ to
first order in the perturbation. It holds that 
$\ffun{GS}A\fvec{GS}=\mathcal C\bffun{GS_L}A_D\bfvec{GS_R}$, where $\mathcal C$ is a normalization
constant such that $1=\mathcal C (GS_L|GS_R)$. Since it already holds that 
$(GS_L|GS_R)=1+\mathcal O\left(\epsilon_\mu^2/G^2\right)$, 
and we are interested in corrections of order
$\epsilon_\mu/G$, we can take $\mathcal C=1$. Thus we have
\begin{widetext}
\begin{eqnarray}
\ffun{GS}A\fvec{GS}&=&\bffun{-}A_{D0}\bfvec{-}+\bffun{GS_L^{(1)}}A_{D+}\Big|-
\Big)+\Big(-\Big|A_{D-}
\bfvec{GS_R^{(1)}}\nonumber\\
&=&2N\overline{A}+\frac{4N(\Omega-N)}{G(\Omega-1)}\left(\overline{\epsilon}\overline{A}
-\frac{\sum_{\mu=-j}^j\epsilon_\mu A_{\mu\mu}}{\Omega}\right),\label{tip14}
\end{eqnarray}
\end{widetext}
where $\overline{A}=\tr A/\Omega$. 

We briefly discuss this novel result: Superconductivity is often presented as the Bose-Einstein condensation
of Cooper pairs. The ground state contains all electrons bound into Cooper pairs with all Cooper pairs
in the same quantum mechanical state. The first term in (\ref{tip14}) confirms this picture: each Cooper
pair contributes $2\overline{A}$ to the expectation value. One could then easily suppose that the kinetic
perturbation breaks up Cooper pairs, and that each Cooper pair can be broken up independently from all
the others, so that the first order correction should also be extensive, i.e. proportional to the number
of Cooper pairs in the ground state. However the first order correction is actually proportional to
$N(\Omega-N)$. 
The fundamental reason for this $N$-dependence is blocking through the exclusion principle. There are only
$2\Omega$ states for the fermions making up the Cooper pairs in the original system to occupy. 
Therefore, when there are $2N$
electrons, all one-particle states are full, and no rearrangement of electrons is possible. In this
case the perturbation has no effect. In the mapped system this ``taking account of the exclusion principle''
manifests itself in the complicated images of the kinetic operator that introduce interactions of ideal
fermions with bosons and with other ideal fermions. 
The deviation from bosonic behavior where the breakup of Cooper
pairs is concerned will be encountered again in the remainder of this text where we will further comment on it.

As a very simple application of (\ref{tip14}), we find the probability $P_\rho$ for the single
particle orbital $\rho$ to be occupied, by calculating the ground-state expectation value of
$\left(a^+_{\rho+}a_{\rho+}+a^+_{\rho-}a_{\rho-}\right)/2$. According to (\ref{tip14}) we 
have
\begin{equation}
P_\rho=\frac{N}{\Omega}+\frac{2N(\Omega-N)}{G(\Omega-1)}\left(\overline{\epsilon}-\epsilon_\rho
\right).\label{tip15}
\end{equation}
As the kinetic term is switched on, single-particle orbitals with below-average kinetic energies
are populated with a higher probability, in order to get a smaller kinetic energy contribution.
In other words, the distribution in (\ref{tip15}) is the result of a competition between the pairing
term, that strives for a distribution where all single particle orbitals are equally likely
to be occupied, and the kinetic term, that would like to populate the $N$ 
one-particle orbitals whose kinetic energies are the lowest. This picture still qualitatively
conforms to our intuition for a realistic superconductor in which the kinetic term cannot
be viewed as a small perturbation to the pairing term. This suggests that insights gained in
an investigation of the strong pairing limit gives qualitative information about superconducting
correlations in a real superconductor.

\section{\label{sect4}Time-dependent Perturbation Theory}
This section is devoted to the perturbative treatment of time-dependent phenomena. We consider
a system that, in the absence of any perturbation, is described by the pairing Hamiltonian in
the seniority limit
$P=-GS^+S$. At a time $t=0$ a perturbation of the form 
\begin{equation}
V=\sum_{\mu,\nu=-j}^j V_{\mu\nu}\left(a_{\mu+}^+a_{\nu+}+a_{\mu-}^+a_{\nu-}\right),\label{tdp1}
\end{equation}
is switched on. 

We will calculate the linear response of the expectation value of an additive one-body operator
$A$ to this perturbation. From standard time-dependent perturbation theory follows that, to
first order in the perturbation
\begin{equation}
\fxpv{A}(t)=\ffun{i}A_I(t)\fvec{i}+\frac{1}{i\hbar}\int_0^tdt'\hspace{1mm}
\ffun{i}\comm{A_I(t),V_I(t')}\fvec{i},\label{tdp2}
\end{equation}
where $A_I(t)=\exp\left(iPt/\hbar\right)A\exp\left(-iPt/\hbar\right)$ and a similar definition holds for $V_I$.
Here $\fvec{i}$ is the state the system was in at time $t=0$, which we will take to be the
unperturbed ground state $\fvec{i}\propto\left(S^+\right)^N\fvec{0}$. Note that if $A$ commutes
with the unperturbed Hamiltonian $P$, and $\fvec{i}$ is a non-degenerate eigenstate of $P$,
then it is also an eigenstate of $A_I(t)$, and hence the expectation value of the commutator in 
(\ref{tdp2}) vanishes. Thus, any additive one-body operator such as the current operator that
is odd under time-reversal and therefore commutes with $P$ (see the discussion following 
(\ref{map12}) in Section \ref{sect1}) has no linear response to an external
perturbation such as $V$.

To compute the matrix 
elements involved in (\ref{tdp2}), we will work with the mapped system. Consider for
instance
\begin{equation}
\ffun{i}A_I(t)V_I(t')\fvec{i}=\bffun{-}A_{ID}(t)V_{ID}(t')\bfvec{-},\label{tdp3}
\end{equation}
where $A_{ID}(t)=\exp\left(iP_Dt/\hbar\right)A_D\exp\left(-iP_Dt/\hbar\right)$ and a similar expression
holds for $V_{ID}(t')$. On the right-hand side of (\ref{tdp3}), we can insert the identity
operator in its spectral representation in the
boson-fermion Hilbert space, between $A_{ID}$ and $V_{ID}$. The 
operators $A_{ID}$ and $V_{ID}$, just like their `Schr\"odinger' analogues, only have 
non-zero matrix elements between eigenstates of the unperturbed Hamiltonian that lie in
the same or adjacent levels. Hence the sum over states in the spectral representation of the
identity operator may be restricted to the levels $s=0$ and $s=1$. As it can easily be seen
that the $s=0$ contributions from the two terms in the commutator cancel we focus on
the $s=1$ terms such as
\begin{equation}
\sum_{\mu\nu=-j}^j\bffun{-}A_{ID}(t)\bfvec{\mu\nu_*}\bffun{\mu\nu_*}V_{ID}(t')\bfvec{-}.
\label{tdp4}
\end{equation}
The states $\bfvec{\mu\nu_*}$ as defined in Section \ref{sect3}
have two ideal fermions, one with spin pointing up and one
with spin pointing down. This is the only possibility we need to include, because the operators
$A$ and $V$ are assumed to be unable to cause spin flipping. This
guarantees that their images, when acting on $\bfvec{-}$ only create ideal fermion 
excitations 
with as many spins pointing up as pointing down. 

Since $\bfvec{-}$ and $\bfvec{\mu\nu_*}$ are
eigenstates of the image of the unperturbed Hamiltonian $P_D$, we have 
\begin{eqnarray}
\bffun{-}A_{ID}(t)\bfvec{\mu\nu_*}&=&\exp\frac{-iGt}{\hbar}\bffun{-}A_{D-}(t)\bfvec{\mu\nu_*},
\label{tdp5}\\
\bffun{\mu\nu_*}A_{ID}(t)\bfvec{-}&=&\exp\frac{iGt}{\hbar}\bffun{\mu\nu_*}A_{D+}(t)\bfvec{-},
\label{tdp6}
\end{eqnarray}
and similarly for the matrix elements of $V_{ID}(t')$, $G$ being the energy difference between the $s=0$
and $s=1$ levels (\ref{ric6}). Thus it holds that
\begin{widetext}
\begin{eqnarray}
\ffun{i}\comm{A_I(t),V_I(t')}\fvec{i}&=&\sum_{\mu\nu=-j}^j
\exp\frac{iG(t'-t)}{\hbar}
\bffun{-}A_{D-}\bfvec{\mu\nu_*}\bffun{\mu\nu_*}V_{D+}\bfvec{-}\nonumber\\
&&\hspace{15mm}-\exp\frac{iG(t-t')}{\hbar}
\bffun{-}V_{D-}\bfvec{\mu\nu_*}\bffun{\mu\nu_*}A_{D+}\bfvec{-}.\label{tdp7}
\end{eqnarray}
At this point the calculation becomes a matter of substituting the correct expressions for the
matrix elements and integrating over $t'$. Finally we find, to first order in $V$
\begin{eqnarray}
\lefteqn{\hspace{-1cm}\fxpv{A}(t)=2N\frac{\tr A}{\Omega}+\frac{4N(\Omega-N)}{G\Omega(\Omega-1)}}\nonumber\\
&&\hspace{1cm}\times\bigg\{\frac{4\tr A\tr V}{\Omega}
-\sum_{\mu\nu=-j}^j\left(A_{\mu\nu}+A_{-\nu-\mu}\right)\left(V_{\nu\mu}
+V_{-\mu-\nu}\right)\bigg\}\sin^2\frac{Gt}{2\hbar}.\label{tdp8}
\end{eqnarray}
\end{widetext}
Note again the $N(\Omega-N)$ factor in front of the first-order correction. As was explained in
the section on time-independent perturbation theory, this factor accounts for exclusion-principle
related deviations from bosonic behaviour by the Cooper pairs. In the context of the perfectly
regular metallic nanograin this has the following implication: Let us suppose that the degeneracy
$2\Omega$ of the partially filled Fermi-level is large compared to the number $N$ of Cooper pairs. 
If we increase the number of Cooper pairs by adding electrons to the system while keeping the
temperature fixed, the linear response to (say)
an external electrical field initially grows, but not linearly. The increase in the linear response
with an increase of particles tapers off as more electrons are added. When the Fermi-level is half-filled
the addition of another Cooper pair does not change the linear response of the system at all. Then,
as more Cooper pairs are added, the linear response actually decreases. 

To see if the result (\ref{tdp8}) makes sense, we consider a one-dimensional system of $2N$ particles on a
ring of length $L$ (with periodic boundary conditions). In this case the index 
$\nu\in-j,-j+1,\dots,j$ is an integer, labeling a one-particle wave function 
$\phi_\mu(x)=\frac{1}{\sqrt{L}}\exp\left(2\pi i\mu x/L\right)$. We will take $V$ to be the
external potential $V(x)=V_0\cos\left(2\pi x/L\right)$, so that $V_{\mu\nu}=V_0\left(
\delta_{\mu,\nu+1}+\delta_{\mu+1,\nu}\right)/2$. We will calculate the linear response of the 
density-of-particles operator 
\begin{equation}
\rho(x)=\sum_{\mu\nu=-j}^j\phi_\mu^*(x)\phi_\nu(x)\left(a_{\mu+}^+a_{\nu+}+a_{\mu-}^+a_{\nu-}\right),
\label{tdp9}
\end{equation}
so that $\rho_{\mu\nu}(x)=\frac{1}{L}\exp\frac{2\pi i(\nu-\mu)x}{L}$. One may ask if it is 
sensible to calculate the linear response of the density of particles. After all, the current
has no linear response. Would the continuity equation $-d\rho/dt=dj/dx$ not imply that the density should then
also have a zero linear response to $V$? It would, but the continuity equation simply does not
hold for the pairing Hamiltonian. (Recall that the continuity equation is usually derived for
a Hamiltonian with a momentum-squared one-body kinetic term and a two-body inter-particle potential
that depends on the distance between pairs of particles only. The pairing Hamiltonian $P$ looks
nothing like this.)

We consider the cut-off
$j$ to have some physical origin, as opposed to it being the result of a regularization 
procedure. For instance, for a large or irregular grain, $2\pi^2j^2\hbar^2/mL^2$ would be set 
equal to the Debye energy of the lattice. For a small regular grain it is
equal to the degeneracy of the partially filled Fermi-level. 
The cut-off is finite and will not be sent to infinity 
at the end of the calculation. Strictly speaking, the operators $\rho$ and $V$ do not leave 
invariant the
fermion space of states with one-particle indices inside the cut-off only. 
For instance, $V$ has, among others, a non-zero 
one-body matrix element $V_{j,j+1}$. It is entirely possible to extend the formalism we have 
developed to accommodate this situation. However, this only leads to small corrections if the
cut-off is large. We therefore simply set the offending matrix elements $V_{\pm j,\pm j\pm1}$
and $V_{\pm j\pm 1,\pm j}$ equal to zero.

We first compute
\begin{eqnarray}
&&\sum_{\mu\nu=-j}^j\left(\rho_{\mu\nu}+\rho_{-\nu-\mu}\right)\left(V_{\nu\mu}+V_{-\mu-\nu}\right)
\nonumber\\
&=&\frac{2V_0}{L}\sum_{\mu\nu=-j}^j\cos\frac{2\pi(\nu-\mu)x}{L}\left(\delta_{\nu,\mu+1}+
\delta_{\mu,\nu+1}\right)\nonumber\\
&=&\frac{4V_0}{L}(\Omega-1)\cos\frac{2\pi x}{L}.\label{tdp10}
\end{eqnarray}
The third line was obtained from the second by noting that there are $2j=\Omega-1$ terms in the
summation over $\mu$ and $\nu$, namely $(\mu,\nu)=(-j+1,-j),(-j+2,-j+1),\ldots,(j,j-1)$ such that
$\delta_{\mu,\nu+1}$ is not zero, and similarly for $\delta_{\mu+1,\nu}$. Noting that $\tr V=0$,
we therefore find that the general formula of (\ref{tdp8}) becomes
\begin{equation}
\fxpv\rho(t)=\frac{2N}{L}-16\frac{NV_0}{LG}\left(1-\frac{N}{\Omega}\right)\cos\frac{2\pi x}{L}
\sin^2\frac{Gt}{2\hbar},\label{tdp11}
\end{equation}
in this specific instance. The spatial factor $-\cos\left(2\pi x/L\right)$ indicates that, as expected,
particles disperse from the potential hill and accumulate in the potential valley. 

As another application of time-dependent perturbation theory, we consider a system that, before
time $t=0$, is described by the Richardson Hamiltonian, but with the pairing constant $G$ so large
that the one-body kinetic term could be ignored. The system is assumed to be in the seniority limit
ground state
$\fvec{GS}\propto\left(S^+\right)^N\fvec{0}$ at $t=0$. Then, at time $t=0$, the pairing interaction
strength is reduced to a value such that the kinetic term has to be taken into account. Such a
tunable pairing interaction might for instance be encountered if our system is an ultra-cold 
trapped atomic Fermi gas. We mention in passing that recently the pairing Hamiltonian $P=-GS^+S$
was discussed in precisely this context.\cite{bgk04} With the aid of the Dyson mapping the
following result is easily obtained. The probability for the system to be found in the 
seniority limit ground state at a time $t$ after the pairing strength was lowered, is to first
non-trivial order in the kinetic energy
\begin{equation}
P_{GS\rightarrow GS}=1-16N\frac{\Omega-N}{\Omega-1}\frac{\overline{\epsilon^2}
-\overline{\epsilon}^2}{G^2}\sin^2\frac{Gt}{2\hbar}.\label{tdp12}
\end{equation}
In this expression, the by now familiar $N(\Omega - N)$ factor instead of simply $N$
again confirms that the likelihood that a Cooper pair will break up into unpaired fermions
depends on the number of availiable unoccupied states for the electrons to scatter into, and
hence on the number of Cooper pairs in the system.

A matter that has not been addressed yet, is how to vary the number of Cooper pairs in the ground state
without varying the number of availible states $\Omega$. While changing the occupancy of the degenerate Fermi 
level in a very small, regular nano-grain is feasible, the ratio $N/\Omega$ in most other systems is not
under experimental control. Rather, the one-particle levels involved in Cooper-pairing lie in a symmetric window
above and below the Fermi-energy. The size of the window is determined by the electron-phonon interaction, and of
the order of the Debye energy, i.e. $\sim 10^{-2}\rm eV$. However, suppose one could find a system where the Fermi-energy
is very close to the bottom or top of the conduction band. We will discuss the case of a Fermi-energy close to the
bottom of the band. This could be attained in an electron doped semi-conductor. By varying the doping concentration
one can shift the Fermi-energy up or down. Thus a situation can be realized where the energy-window for levels to
participate in Cooper-pairing extends below the conduction band. There will then be fewer occupied levels than unoccupied
levels in the energy-window, simply because there are no states in the section of the window that extends below the conduction 
band. The ratio $N/\Omega$ will be less than a half, and furthermore, it is tunable: by varying
the doping concentration, the number of electrons in the conduction band, and hence the Fermi-energy and the number of
states within the relevant energy window can be increased or decreased. Of course, one needs to find a semi-conductor
in which the lack of screening of the electron-electron interaction does not override the pairing interaction.

\section{\label{summ}Summary}
The salient points in this paper are the following. We used the
Dyson mapping as a similarity transform to map the Richardson Hamiltonian onto an equivalent 
boson-fermion Hamiltonian. This Hamiltonian uses bosons to describe
the bound fermion pairs that result from the pairing interaction, instead of the time-reversed collective 
fermion pairs created by the operator $S^+$. From the structure of the 
mapped operator emerges our first main result: An additive one-body operator only has 
non-zero matrix elements between eigenstates of the seniority pairing model, if these eigenstates
come from the same or adjacent energy levels in the spectrum of the pairing Hamiltonian. 
With this selection rule it becomes feasible to investigate the strong pairing limit of the
Richardson Hamiltonian without relying on the Bethe-Ansatz solvability of the Richardson
Hamiltonian. While our aim remains the same as that of previous authors, i.e. to develop
a simple and intuitive analytical picture of the low-temperature physics of superconducting
metallic nano-grains, the new tools at our disposal allow us to obtain new results:
\begin{enumerate}
\item{In Section \ref{sect3} we calculated the ground state expectation value of an arbitrary
one-body operator for a system of $2N$ electrons in the strong pairing limit. 
In Section \ref{sect4} we showed that the selection rule we identified and used in conjunction with the 
Dyson mapping also facilitates
time-dependent perturbation theory. Here we explicitly perturbed the strong pairing limit
with a more general perturbation than the kinetic term in the Richardson Hamiltonian. Results
obtained in both calculations show that the likelihood for a perturbation
to break up a Cooper pair depends on the number density of Cooper pairs in the system. This 
is explained by the fact that the electrons that constitute a Cooper pair 
obey the exclusion principle and underlines that there are circumstances in which
Cooper pairs do not behave as bosons.}
   
\item{Also in Section \ref{sect3} we calculated the occupation probability of single particle
levels in the strong pairing limit. We found that a picture emerges of a competition between
the kinetic term that favours a Fermi-sea, and the pairing term that favours a situation
where the occupation probabilities of all one-particle levels are the same. This picture
agrees qualitatively with our intutition also for a real superconductor in which the kinetic
term cannot be treated as a perturbation. Since there is a smooth transition from the stong
pairing limit to more realistic values for the pairing interaction strength, we expect that
results such as the non-extensivity of the linear response also holds for the nanograins 
studied experimentally.}
\end{enumerate}
\begin{acknowledgments}
We are indebted to Hannes Kriel for his careful proof reading of the final manuscript.
One of us (IS) thanks the National Research Foundation and the Harry Crossley Foundation for
funding received while conducting this research. This work is supported through grant GUN2053779
of the NRF. HBG thanks the Institute for Nuclear Theory at the University of Washington for its partial support, 
hospitality and conducive working conditions during completion of this work.

\end{acknowledgments}
\appendix*
\section{\label{appA}Degenerate Perturbation Theory}
In this appendix the general formalism is presented in detail, for finding perturbatively
the spectra of (possibly non-hermitian) operators if there is degeneracy when the perturbation
is switched off. The argument is developed along the lines of the non-degenerate (hermitian) case 
as treated by Sakurai.\cite{sak94} 
We could not locate any source that gives a satisfactory treatment of the degenerate case.

We are given an operator $H(\lambda)=H_0+\lambda H_1$. Neither $H_0$ nor $H_1$ have to be 
hermitian. 
We assume that $H$ is fully diagonalizable, at least in a finite region around $\lambda=0$. For 
$\lambda\not=0$, we assume that $H(\lambda)$ has a non-degenerate spectrum, so that a single 
label $\alpha=1,2,\ldots,M$ may be used to specify eigenvalues and eigenvectors uniquely:
\begin{equation}
H(\lambda)\rvec{\alpha}_\lambda=E_\alpha(\lambda)\rvec{\alpha}_\lambda,\label{tipert1}
\end{equation}
with $E_\alpha(\lambda)\not=E_\beta(\lambda)$ if $\alpha\not=\beta$ and $\lambda\not=0$.
The subscript $R$ is used to indicate that we are dealing with right-eigenvectors, later to be 
contrasted
with left-eigenvectors, that do not coincide with the right eigenvectors when $H(\lambda)$ is not
hermitian.
We take it that all the $E_\alpha(\lambda)$ and $\rvec{\alpha}_\lambda$ are analytical functions
of $\lambda$ in a finite region around $\lambda=0$. Furthermore, if two eigenvalues converge at 
$\lambda=0$, i.e. $E_\alpha(0)=E_\beta(0)$, we assume that these eigenvalues have different
first derivatives at $\lambda=0$. This condition is referred to by saying that all degeneracy is
lifted in the first order.
At $\lambda=0$, the spectrum of $H$ may be degenerate. 
Our notation takes care of this as follows:
At $\lambda=0$ we let $M'$ be the number
of distinct eigenvalues of $H_0$. We then indicate these distinct eigenvalues of $H_0$ as 
$\En_l$, $l=1,2,\ldots,M'$. 
We partition the labels $\alpha=1,2,\ldots,M$ into disjoint sets $P_l$, labeled by integers
$l=1,2,\ldots M'$, by defining $P_l$ to be the set of all indices $\alpha$ such 
that $E_\alpha$ flows
to $\En_l$ as $\lambda$ goes to zero. We can then also define a function $p$ from the index set
$1,2,\ldots,M$ to the index set $1,2,\ldots,M'$ as follows: $p(\alpha)=l$ where $l$ is the
unique label such that $\alpha\in P_l$. We refer to the space spanned by all eigenvectors of $H_0$
with eigenvalue $\En_l$ as the $\En_l$ eigenspace of $H_0$. It is assumed we know a
basis for each $\En_l$ eigenspace of $H_0$. The basis elements that span the $\En_l$ eigenspace
will be denoted $\left|\psi,l_R\right)$ where $\psi$ is a discrete index that runs from $1$ to
some integer $\omega_l$ which is the dimension of the $\En_l$ eigenspace. 
The set $\left\{\left|\psi,l_R\right):l=1,\ldots,M';\psi=1,\ldots,\omega_l\right\}$
forms a basis for the domain of $H(\lambda)$. 
It is in terms of
these basis states that we want to express the eigenstates of $H(\lambda)$ approximately.
Corresponding to this basis we can always find a left basis 
$\left\{\left|\psi,l_L\right):l=1,\ldots,M';\psi=1,\ldots,\omega_l\right\}$
such that the equations
\begin{equation}
\left(\psi,l_L\right|\left.\phi,m_R\right)=\delta_{l,m}\delta_{\psi,\phi}, \label{tipert3c}
\end{equation}
hold for all $l,m,\psi$ and $\phi$. 

We uniquely define a projection operator $\Pi_l$ as the linear operator that maps any vector in the 
$\En_l$ eigenspace of $H_0$ onto itself, while mapping any vector that can be written as 
a linear combination of vectors from the other eigenspaces of $H_0$, onto zero. (This might not
be an {\em orthogonal} projection operator though. When $H_0$ is non-hermitian, there are $\En_l$ eigenspaces 
that are not orthogonal to each other.)
An operator $\tilde{\Pi}_l$ is defined as $\tilde{\Pi}_l={\rm I}-\Pi_l$. In terms of the states
$\left|\psi,l_L\right)$ and $\left|\phi,m_R\right)$, the identity operator ${\rm I}$, the 
unperturbed operator $H_0$ and the projection operators $\Pi_l$ and $\tilde{\Pi}_l$ can be 
expressed as follows:
\begin{eqnarray}
{\rm I}&=&\sum_{l=1}^{M'}\sum_{\psi=1}^{\omega_l}\left|\psi,l_R\right)\left(\psi,l_L\right|,
\label{tipert4a}\\
{H_0}&=&\sum_{l=1}^{M'}\En_l\sum_{\psi=1}^{\omega_l}\left|\psi,l_R\right)\left(\psi,l_L\right|,
\label{tipert4b}\\
\Pi_l&=&\sum_{\psi=1}^{\omega_l}\left|\psi,l_R\right)\left(\psi,l_L\right|,
\label{tipert4}\\
\tilde{\Pi}_l&=&\sum_{k=1\not=l}^{M'}\sum_{\psi=1}^{\omega_k}\left|\psi,k_R\right)
\left(\psi,k_L\right|,
\label{tipert5}
\end{eqnarray}
Note that $\Pi_l\Pi_m=\delta_{lm}\rm \Pi_m$.

With the above preliminaries out of the way we now expand the eigenvalues 
$E_\alpha(\lambda)$ and
eigenvectors $\rvec{\alpha}_\lambda$ of $H(\lambda)$ in terms of $\lambda$:
\begin{eqnarray}
E_\alpha(\lambda)&=&\En_{p(\alpha)}+\lambda\Delta_\alpha^{(1)}+\lambda^2\Delta_\alpha^{(2)}
+\ldots\nonumber\\
\rvec{\alpha}_\lambda&=&\rvec{\alpha^{(0)}}+\lambda\rvec{\alpha^{(1)}}+\lambda^2\rvec{\alpha^{(2)}}
+\ldots\label{tipert2}
\end{eqnarray}
This definition fixes the directions of the $\lambda$-independent zero'th order eigenvectors
$\rvec{\alpha^{(0)}}_0$ uniquely as the direction to which the $\alpha$-eigenvector of $H(\lambda)$ 
converges as $\lambda$ goes to zero. This direction is well-defined since we assumed that 
$H(\lambda)$ has a non-degenerate spectrum for $\lambda\not=0$. The magnitudes of the eigenvectors
are left arbitrary.

The set of eigenvectors $\left\{\rvec{\alpha^{(0)}}\right\}_{\alpha=1}^{M}$ also forms a 
basis for
the domain of $H(\lambda)$. We use it to define a second basis 
(sometimes called the contra-variant basis) $\left\{\lvec{\alpha^{(0)}}\right\}_{\alpha=1}^M$
through the set of equations
\begin{equation}
\lfunc{\alpha}\left.\beta^{(0)}_R\right)=\delta_{\alpha,\beta}.\label{tipert3}
\end{equation}
Equation (\ref{tipert3}) implies that $\lfunc{\alpha}H_0=\En_{p(\alpha)}\lfunc{\alpha}$.
Using the states $\left\{\rvec{\alpha^{(0)}}\right\}_{\alpha=1}^{M}$ and functionals
$\left\{\left(\alpha_L^{(0)}\right|\right\}_{\alpha=1}^{M}$ we define another set of projection
operators that will come in handy during our calculations. Define a projection operator
\begin{equation}
\pi_\alpha=\left|\alpha_R^{(0)}\right)\left(\alpha_L^{(0)}\right|,\label{tipert4x}
\end{equation}
that singles out the $\left|\alpha_R^{(0)}\right)$ component of the decomposition of any vector in 
the $\left\{\rvec{\alpha^{(0)}}\right\}_{\alpha=1}^{M}$ basis. Let $\tilde{\pi}_\alpha$ be the
complementary projection operator of $\pi_\alpha$ in the $\mathcal E_{p(\alpha)}$ subspace by 
setting 
\begin{equation}
\tilde{\pi}_\alpha=\sum_{\beta\in P_{p(\alpha)}\setminus\{\alpha\}}\left|\beta_R^{(0)}\right)
\left(\beta_L^{(0)}\right|=\Pi_{p(\alpha)}-\pi_\alpha.\label{tipert5xx}
\end{equation}
Thus, the identity operator may be decomposed as
\begin{equation}
{\rm I}=\tilde{\Pi}_{p(\alpha)}+\tilde{\pi}_\alpha+\pi_\alpha.\label{tipert5x}
\end{equation}
The next step is to choose the relative normalization between the eigenstate 
$\left|\alpha_R\right)_\lambda$ and $\left|\alpha_L\right)_\lambda$. 
The most convenient choice turns out to be 
$\Big(\alpha_L^{(0)}\Big|\alpha_R\Big)_\lambda=1$. With this choice it follows that
\begin{equation}
\pi_\alpha\left|\alpha_R\right)_\lambda=\left|\alpha_R^{(0)}\right).\label{tipert6x}
\end{equation}
Now we rewrite the eigenvalue equation in the form
\begin{equation}
\left(\mathcal E_{p(\alpha)}-H_0\right)\left|\alpha_R\right)_\lambda=\left(\lambda H_1-
\Delta_\alpha(\lambda)\right)\left|\alpha_R\right)_\lambda,\label{tipert7x}
\end{equation}
where $\Delta_{\alpha}(\lambda)=\sum_{k=1}^\infty\Delta_\alpha^{(k)}\lambda^k$. 

The first thing
we do with this equation is multiply it by $\Pi_{p(\alpha)}$, noting that  
$\Pi_{p(\alpha)}\left(\mathcal E_{p(\alpha)}-H_0\right)=0$, to find
\begin{equation}
\Pi_{p(\alpha)}\left(\lambda H_1-\Delta_\alpha(\lambda)\right)\left|\alpha_R\right)_\lambda=0,
\label{tipert8x}
\end{equation}
which has to hold order for order in $\lambda$.
Note that, by definition, the zero'th order term in the $\lambda$-expansion of 
$\left|\alpha_R\right)_\lambda$, namely $\left|\alpha_R^{(0)}\right)$, is an element of the 
$\mathcal E_{p(\alpha)}$ eigenspace of $H_0$ and hence 
$\Pi_{p(\alpha)}\left|\alpha_R^{(0)}\right)=\left|\alpha_R^{(0)}\right)$. Keeping this in mind, we
look at the terms in (\ref{tipert8x}) of lowest order in $\lambda$ to find
\begin{equation}
\Pi_{p(\alpha)}H_1\left|\alpha_R^{(0)}\right)=\Delta_\alpha^{(1)}\left|\alpha_R^{(0)}\right).
\label{tipert9x}
\end{equation}
This eigenvalue equation enables us to find the zero'th order vectors $\left|\alpha_R^{(0)}\right)$
as the eigenvectors of $\Pi_{p(\alpha)}H_1$ in the $\mathcal E_{p(\alpha)}$ eigenspace. Since we
assume all degeneracy to be lifted in the first order, no two eigenvalues $\Delta_\alpha^{(1)}$
and $\Delta_\beta^{(1)}$ with $\alpha$, $\beta\in P_{p(\alpha)}$ are the same, and 
(\ref{tipert9x}) is necessary and sufficient to determine the various $\left|\alpha_R^{(0)}\right)$
in the $\mathcal E_{p(\alpha)}$ eigenspace.

In order to proceed, we define an operator
\begin{equation}
Q_k=\sum_{l\not=k}\frac{1}{\mathcal E_k-\mathcal E_l}\Pi_l.\label{tipert10x}
\end{equation}
We return to (\ref{tipert7x}) and multiply it with $Q_{p(\alpha)}$, noting that 
$Q_{p(\alpha)}\left(\mathcal E_{p(\alpha)}-H_0\right)=\tilde{\Pi}_{p(\alpha)}.$ Thus we arrive at
one of the most important formulas in the present discussion, namely
\begin{equation}
\tilde{\Pi}_{p(\alpha)}\left|\alpha_R\right)_\lambda=Q_{p(\alpha)}\left(\lambda H_1
-\Delta_\alpha(\lambda)\right)\left|\alpha_R\right)_\lambda.\label{tipert11x}
\end{equation}
This equation has, on the left-hand side, the part of the eigenvector 
$\left|\alpha_R\right)_\lambda$ that can be written as a linear combination of vectors 
in $\mathcal E_l$ eigenspaces with $l\not=p(\alpha)$. We want a similar equation for that part
of $\left|\alpha_R\right)_\lambda$ that lies inside the $\mathcal E_{p(\alpha)}$ eigenspace.

For this purpose, we firstly rewrite (\ref{tipert8x}) to read
\begin{eqnarray}
\lefteqn{\hspace{-7mm}\left(\Delta_\alpha(\lambda)-\lambda\Pi_{p(\alpha)}H_1\Pi_{p(\alpha)}\right)\Pi_{p(\alpha)}
\left|\alpha_R\right)_\lambda}\nonumber\\
&&{}\hspace{7mm}-\lambda\Pi_{p(\alpha)}H_1\tilde{\Pi}_{p(\alpha)}\left|\alpha_R\right)
_\lambda=0,\label{tipert12x}
\end{eqnarray}
by recalling that $\Pi_{p(\alpha)}+\tilde{\Pi}_{p(\alpha)}={\rm I}$ and 
$\left(\Pi_{p(\alpha)}\right)^2=\Pi_{p(\alpha)}$. Now we substitute (\ref{tipert11x}) into the
second term of (\ref{tipert12x}) to find
\begin{eqnarray}
\lefteqn{\hspace{-2mm}\left(\Delta_\alpha(\lambda)-\lambda\Pi_{p(\alpha)}H_1\Pi_{p(\alpha)}\right)\Pi_{p(\alpha)}
\left|\alpha_R\right)_\lambda}\nonumber\\
&&{}\hspace{2mm}-\lambda\Pi_{p(\alpha)}H_1Q_{p(\alpha)}\left(\lambda H_1
-\Delta_\alpha(\lambda)\right)\left|\alpha_R\right)_\lambda=0.\nonumber\\
\label{tipert13x}
\end{eqnarray}
Define an operator 
\begin{equation}
q_\alpha=\sum_{\beta\in P_{p(\alpha)}\setminus\{\alpha\}}\frac{1}{\Delta_\alpha^{(1)}-
\Delta_\beta^{(1)}}\left|\beta_R^{(0)}\right)\left(\beta_R^{(0)}\right|.\label{tipert14x}
\end{equation}
Note that $q_\alpha\left(\Delta_\alpha^{(1)}-\Pi_{p(\alpha)}H_1\Pi_{p(\alpha)}\right)=
\tilde{\pi}_\alpha$ and that $q_\alpha\Pi_{p(\alpha)}=q_\alpha$. Hence, by multiplying  
(\ref{tipert13x}) with $q_\alpha$ and dividing by $\lambda$, note that
\begin{eqnarray}
\samepage
\tilde{\pi}_\alpha\left|\alpha_R\right)_\lambda&=&q_\alpha H_1Q_{p(\alpha)}\left(\lambda H_1-
\Delta_\alpha(\lambda)\right)\left|\alpha_R\right)_\lambda\nonumber\\
&&\hspace{2mm}-\sum_{k=1}^\infty \Delta_\alpha^{(k+1)}
\lambda^k q_\alpha\left|\alpha_R\right)_\lambda.
\label{tipert15x}
\end{eqnarray}
This is the counterpart of (\ref{tipert11x}) that we needed for the part of 
$\left|\alpha_R\right)_\lambda$ that lies inside the $\mathcal E_{p(\alpha)}$ eigenspace. When
we combine the results of (\ref{tipert6x}), (\ref{tipert11x}) and (\ref{tipert15x}), 
remembering that the identity operator may be decomposed as ${\rm I}=\tilde{\Pi}_{p(\alpha)}
+\tilde{\pi}_\alpha+\pi_\alpha$ we find
\begin{widetext}
\begin{equation}
\left|\alpha_R\right)_\lambda=\left|\alpha_R^{(0)}\right)+\left(1+q_\alpha H_1\right) Q_{p(\alpha)}
\left(\lambda H_1-\Delta_\alpha(\lambda)\right)\left|\alpha_R\right)_\lambda-\sum_{k=1}^\infty
\Delta_{\alpha}^{(k+1)}\lambda^k q_\alpha\left|\alpha_R\right)_\lambda.\label{tipert16x}
\end{equation}
Collecting terms of order $N+1$ we find, for $N\geq0$,
\begin{eqnarray}
\left|\alpha_R^{(N+1)}\right)&=&\left(1+q_\alpha H_1\right)Q_{p(\alpha)}\left\{H_1\left|
\alpha_R^{(N)}\right)-\sum_{M=1}^{N+1}\Delta_{\alpha}^{(M)}\left|\alpha_R^{(N+1-M)}\right)\right\}
\nonumber\\
&&\hspace{2mm}-\sum_{M=2}^{N+2}\Delta_\alpha^{(M)}q_\alpha\left|\alpha_R^{(N+2-M)}\right).
\label{tipert17x}
\end{eqnarray}
\end{widetext}
Note that $Q_{p(\alpha)}\left|\alpha_R^{(0)}\right)=0$ and $q_\alpha\left|\alpha_R^{(0)}\right)=0$.
Thus, for $N\geq1$, the upper bounds on the two summations in the above expression may respectively 
be decreased from $N+1$ to $N$ and from $N+2$ to $N+1$, while for the case where $N=0$, the two 
summations may be left out entirely. To complete the expansions, we need an expression for 
$\Delta_\alpha^{(M)}$. This we find by multiplying (\ref{tipert7x}) with the functional 
$\left(\alpha_R^{(0)}\right|$, recalling our normalization convention 
$\Big(\alpha_L^{(0)}\Big|\alpha_R\Big)_\lambda=1$, and collecting terms of order $M$ in $\lambda$
to find
\begin{equation}
\Delta_\alpha^{(M)}=\left(\alpha_L^{(0)}\right|H_1\left|\alpha_R^{(M-1)}\right).\label{tipert18x}
\end{equation}
Our final results are then:
\begin{widetext}
\begin{eqnarray}
\left|\alpha_R^{(N+1)}\right)&=&\left(1+q_\alpha H_1\right)Q_{p(\alpha)}\left\{H_1\left|
\alpha_R^{(N)}\right)-\sum_{M=1}^{N}\Delta_{\alpha}^{(M)}\left|\alpha_R^{(N+1-M)}\right)\right\}\nonumber\\
&&\hspace{2mm}-\sum_{M=2}^{N+1}\Delta_\alpha^{(M)}q_\alpha\left|\alpha_R^{(N+2-M)}\right),
\label{tipert19x}
\end{eqnarray}
\end{widetext}
and
\begin{equation}
\Delta_\alpha^{(N+1)}=\left(\alpha_L^{(0)}\right|H_1\left|\alpha_R^{(N)}\right).\label{tipert20x}
\end{equation}
It is understood that, in the case where $N=0$ and the lower bounds of the summations in
(\ref{tipert19x}) exceed the upper bounds, the summation contains no terms, i.e. 
$\left|\alpha_R^{(1)}\right)=\left(1+q_\alpha H_1\right)Q_{p(\alpha)}H_1\left|\alpha_R^{(0)}
\right)$. The formulas of (\ref{tipert19x}) and (\ref{tipert20x}) are all we need to calculate
the corrections of order $N+1$ to both the eigenvectors and eigenvalues of $H$, if the corrections
of orders $0,1,2,\ldots,N$ are known. Together with the initial conditions provided by the 
eigenvalue equation
\begin{equation}
\Pi_{p(\alpha)}H_1\left|\alpha_R^{(0)}\right)=\Delta_\alpha^{(1)}\left|\alpha_R^{(0)}\right)
\label{tipert21x}
\end{equation}
this allows us to write down in principal the expansion of any eigenvector and its eigenvalue to
arbitrary order in $\lambda$. Concerning (\ref{tipert20x}) for the eigenvalue-correction of order 
$N+1$, note that $\left(\alpha_L^{(0)}\right|H_1q_\alpha=0$. This is true because the states
$\left|\alpha_R^{(0)}\right)$ and functionals $\left(\beta_L^{(0)}\right|$ with $\alpha$, 
$\beta\in P_l$ were chosen to diagonalize $H_1$ in the $\mathcal E_{p(\alpha)}$ eigenspace, 
in the sense that
$\left(\beta_L^{(0)}\right|H_1\left|\alpha_R^{(0)}\right)=\Delta_\alpha^{(1)}
\delta_{\alpha,\beta}$. This means that, when we substitute from (\ref{tipert19x}) for
$\left|\alpha_R^{(N)}\right)$ in (\ref{tipert20x}), the terms that are multiplied from
the left with an operator $q_\alpha$ disappear and we are left with
\begin{eqnarray}
\lefteqn{\Delta_\alpha^{(N+2)}=\left(\alpha_L^{(0)}\right|H_1Q_{p(\alpha)}H_1\left|
\alpha_R^{(N)}\right)}\nonumber\\
&&-\sum_{M=1}^{N}\Delta_{\alpha}^{(M)}
\left(\alpha_L^{(0)}\right|H_1Q_{p(\alpha)}\left|\alpha_R^{(N+1-M)}\right),\nonumber\\
\label{tipert22x}
\end{eqnarray}
which is an expression for the eigenvalue-correction, with all the dead-wood cut away.
If an unperturbed eigenstate $\left|\alpha_R^{(0)}\right)$ has a non-degenerate eigenvalue,
i.e. $H_0$ does not contain other eigenvectors with the same eigenvalue as 
$\left|\alpha_R^{(0)}\right)$, the situation and hence the formulas are simpler than above. In
this case, the proper operator to use for $q_\alpha$ is simply the zero-operator. With this
adjustment the general formulas reduce to formulas valid for the specific case of no degeneracy.

Below, we write down explicitly the first order correction to an eigenstate and the second
order correction to its eigenvalue, for the general case where degeneracy is present in the 
unperturbed system:
\begin{eqnarray}
\lefteqn{
\left|\alpha_R^{(1)}\right)=\sum_{l\not=p(\alpha)}\frac{\Pi_lH_1\left|\alpha_R^{(0)}\right)}
{\mathcal E_{p(\alpha)}-\mathcal E_l}}\nonumber\\
&&+\sum_{l\not=p(\alpha)}\hspace{2mm}\sum_{\beta\in P_{p(\alpha)}
\setminus\{\alpha\}}\frac{\left(\beta_L^{(0)}\right|H_1\Pi_lH_1\left|\alpha_R^{(0)}\right)
\left|\beta^{(0)}_R\right)}{\left(\mathcal E_{p(\alpha)}-\mathcal E_l\right)
\left(\Delta_\alpha^{(1)}-\Delta_\beta^{(1)}\right)},\nonumber\\
\label{tipert23x}
\end{eqnarray}
\begin{equation}
\Delta^{(2)}_\alpha=\sum_{l\not=p(\alpha)}\frac{\lfunc{\alpha}H_1\Pi_l
H_1\rvec{\alpha^{(0)}}}{\En_{p(\alpha)}-\En_l}\label{tipert24x}
\end{equation}
Note that the formula for the second order correction to the eigenvalue, (\ref{tipert24x})
looks the same for the degenerate case as for the non-degenerate case. The only difference is
that, in the degenerate case, the zero'th order eigenvector $\left|\alpha_R^{(0)}\right)$ is
a simultaneous eigenvector of $H_0$ and $\Pi_{p(\alpha)}H_1$, lying in the 
$\mathcal E_{p(\alpha)}$ eigenspace, whereas in the non-degenerate case it is uniquely fixed
by diagonalizing $H_0$.
 
When we apply this perturbation expansion method in conjunction with the Dyson
mapping, there is still the matter of the physical subspace to address.
The question is: If we approximate the eigenvectors of $H(\lambda)$ by including only a few terms
in the perturbation expansion, do we lose the notion of physical eigenstates that lie exactly
in the physical subspace? Recall that if we diagonalize $H(\lambda)$ exactly, we can express the
physical subspace as the space spanned by a certain subset of the eigenvectors of $H(\lambda)$.
If we only approximate the eigenvectors by the first few terms of their $\lambda$ expansions, 
can we still find a subset of these approximate eigenvectors that span the physical subspace? 
Or do the $\lambda$-expansions for the
physical eigenvectors, if truncated after a few terms, only lie close to, but not necessarily in
the physical subspace? Perhaps not unexpectedly, the answer is that the
approximate physical eigenvectors still lie exactly in the physical subspace. This can be seen 
as follows. Any operator that leaves the physical subspace invariant is said to be
a physical operator. Then if we assume that $H(\lambda)$ is the linear extension of the
image of a fermion operator for all 
$\lambda$, it follows that $H_0$ and $H_1$ are both physical operators. Furthermore, it
is a simple matter to show that the operators $\Pi_l$ and $Q_l$
are physical operators. This means that $\Pi_l H_1$ 
is also a physical operator. We find the zero'th order eigenvectors $\rvec{\alpha^{(0)}}$ by
diagonalizing the operator $\Pi_{p(\alpha)} H_1$ in the $\En_{p(\alpha)}$ eigenspace of $H_0$.
Since $\Pi_{p(\alpha)} H_1$ is a physical operator with a non-degenerate spectrum, the eigenvectors
thus obtained can be separated into a set that spans the overlap of physical subspace
and the $\En_{p(\alpha)}$ eigenspace, and a set of ghost states. Since the operator $q_\alpha$
is defined in such a way that it has the same eigenvectors in the 
$\mathcal E_{p(\alpha)}$ eigenspace as the physical operator $\Pi_{p(\alpha)}H_1$, it follows
that $q_\alpha$ is a physical operator. Then, if we start the recursion
of (\ref{tipert10x}) with a state from the physical sector, the corrections that the recursion
generates remain inside the physical subspace. This follows inductively from the fact that
higher order corrections in (\ref{tipert19x}) are generated by acting with physical operators
on lower corrections. If the zero'th order eigenvector is physical, this implies that the first
order correction is physical. If the zero'th order eigenvector and the first order correction are
both physical, then so too is the second order correction, etc.
This implies that the exact physical eigenstates, when expanded in terms of $\lambda$, consist
of terms that all lie in the physical subspace themselves. Truncating the expansions after a
finite number of terms therefore still leaves one in the physical subspace.


\begin{thebibliography}{26}
\expandafter\ifx\csname natexlab\endcsname\relax\def\natexlab#1{#1}\fi
\expandafter\ifx\csname bibnamefont\endcsname\relax
  \def\bibnamefont#1{#1}\fi
\expandafter\ifx\csname bibfnamefont\endcsname\relax
  \def\bibfnamefont#1{#1}\fi
\expandafter\ifx\csname citenamefont\endcsname\relax
  \def\citenamefont#1{#1}\fi
\expandafter\ifx\csname url\endcsname\relax
  \def\url#1{\texttt{#1}}\fi
\expandafter\ifx\csname urlprefix\endcsname\relax\def\urlprefix{URL }\fi
\providecommand{\bibinfo}[2]{#2}
\providecommand{\eprint}[2][]{\url{#2}}

\bibitem[{\citenamefont{von Delft and Ralph}(2001)}]{vdr01}
\bibinfo{author}{\bibfnamefont{J.}~\bibnamefont{von Delft}} \bibnamefont{and}
  \bibinfo{author}{\bibfnamefont{D.~C.} \bibnamefont{Ralph}},
  \bibinfo{journal}{Phys.\ Rep.} \textbf{\bibinfo{volume}{345}},
  \bibinfo{pages}{61} (\bibinfo{year}{2001}).

\bibitem[{\citenamefont{Kurland et~al.}(2000)\citenamefont{Kurland, Aleiner,
  and Altshuler}}]{kaa00}
\bibinfo{author}{\bibfnamefont{I.~L.} \bibnamefont{Kurland}},
  \bibinfo{author}{\bibfnamefont{I.~L.} \bibnamefont{Aleiner}},
  \bibnamefont{and} \bibinfo{author}{\bibfnamefont{B.~L.}
  \bibnamefont{Altshuler}}, \bibinfo{journal}{Phys.\ Rev.\ B}
  \textbf{\bibinfo{volume}{62}}, \bibinfo{pages}{14886} (\bibinfo{year}{2000}).

\bibitem[{\citenamefont{Richardson}(1963{\natexlab{a}})}]{ric63a}
\bibinfo{author}{\bibfnamefont{R.~W.} \bibnamefont{Richardson}},
  \bibinfo{journal}{Phys.\ Lett.} \textbf{\bibinfo{volume}{3}},
  \bibinfo{pages}{277} (\bibinfo{year}{1963}{\natexlab{a}}).

\bibitem[{\citenamefont{Richardson}(1963{\natexlab{b}})}]{ric63b}
\bibinfo{author}{\bibfnamefont{R.~W.} \bibnamefont{Richardson}},
  \bibinfo{journal}{Phys.\ Lett.} \textbf{\bibinfo{volume}{5}},
  \bibinfo{pages}{82} (\bibinfo{year}{1963}{\natexlab{b}}).

\bibitem[{\citenamefont{Yuzbashyan et~al.}(2003)\citenamefont{Yuzbashyan,
  Baytin, and Altshuler}}]{yba03}
\bibinfo{author}{\bibfnamefont{E.~A.} \bibnamefont{Yuzbashyan}},
  \bibinfo{author}{\bibfnamefont{A.~A.} \bibnamefont{Baytin}},
  \bibnamefont{and} \bibinfo{author}{\bibfnamefont{B.~L.}
  \bibnamefont{Altshuler}}, \bibinfo{journal}{Phys.\ Rev.\ B}
  \textbf{\bibinfo{volume}{68}}, \bibinfo{pages}{214509}
  (\bibinfo{year}{2003}).

\bibitem[{\citenamefont{Dobaczewski et~al.}(1993)\citenamefont{Dobaczewski,
  Scholtz, and Geyer}}]{dsg93}
\bibinfo{author}{\bibfnamefont{J.}~\bibnamefont{Dobaczewski}},
  \bibinfo{author}{\bibfnamefont{F.~G.} \bibnamefont{Scholtz}},
  \bibnamefont{and} \bibinfo{author}{\bibfnamefont{H.~B.} \bibnamefont{Geyer}},
  \bibinfo{journal}{Phys.\ Rev.\ C} \textbf{\bibinfo{volume}{48}},
  \bibinfo{pages}{2313} (\bibinfo{year}{1993}).

\bibitem[{\citenamefont{Navratil et~al.}()\citenamefont{Navratil, Geyer, and
  Dobaczewski}}]{ngd94}
\bibinfo{author}{\bibfnamefont{P.}~\bibnamefont{Navratil}},
  \bibinfo{author}{\bibfnamefont{H.~B.} \bibnamefont{Geyer}}, \bibnamefont{and}
  \bibinfo{author}{\bibfnamefont{J.}~\bibnamefont{Dobaczewski}},
  \eprint{nucl-th/9405021}.

\bibitem[{\citenamefont{Navratil et~al.}(1995)\citenamefont{Navratil, Geyer,
  and Dobaczewski}}]{ngd95}
\bibinfo{author}{\bibfnamefont{P.}~\bibnamefont{Navratil}},
  \bibinfo{author}{\bibfnamefont{H.~B.} \bibnamefont{Geyer}}, \bibnamefont{and}
  \bibinfo{author}{\bibfnamefont{J.}~\bibnamefont{Dobaczewski}},
  \bibinfo{journal}{Phys.\ Rev.\ C} \textbf{\bibinfo{volume}{52}},
  \bibinfo{pages}{1394} (\bibinfo{year}{1995}).

\bibitem[{\citenamefont{Navratil et~al.}(1996)\citenamefont{Navratil, Geyer,
  and Dobaczewski}}]{ngd96}
\bibinfo{author}{\bibfnamefont{P.}~\bibnamefont{Navratil}},
  \bibinfo{author}{\bibfnamefont{H.~B.} \bibnamefont{Geyer}}, \bibnamefont{and}
  \bibinfo{author}{\bibfnamefont{J.}~\bibnamefont{Dobaczewski}},
  \bibinfo{journal}{Nucl.\ Phys.\ A} \textbf{\bibinfo{volume}{607}},
  \bibinfo{pages}{23} (\bibinfo{year}{1996}).

\bibitem[{\citenamefont{Cejnar and Geyer}(2002)}]{cg02}
\bibinfo{author}{\bibfnamefont{P.}~\bibnamefont{Cejnar}} \bibnamefont{and}
  \bibinfo{author}{\bibfnamefont{H.~B.} \bibnamefont{Geyer}},
  \bibinfo{journal}{Phys.\ Rev.\ C} \textbf{\bibinfo{volume}{65}},
  \bibinfo{pages}{044313} (\bibinfo{year}{2002}).

\bibitem[{\citenamefont{Ring and Schuck}(1980)}]{rs80}
\bibinfo{author}{\bibfnamefont{P.}~\bibnamefont{Ring}} \bibnamefont{and}
  \bibinfo{author}{\bibfnamefont{P.}~\bibnamefont{Schuck}},
  \emph{\bibinfo{title}{The Nuclear Many-Body Problem}}
  (\bibinfo{publisher}{Springer-Verlag}, \bibinfo{address}{New York},
  \bibinfo{year}{1980}), \bibinfo{note}{chap. 9}.

\bibitem[{\citenamefont{Klein and Marshalek}(1991)}]{km91}
\bibinfo{author}{\bibfnamefont{A.}~\bibnamefont{Klein}} \bibnamefont{and}
  \bibinfo{author}{\bibfnamefont{E.~R.} \bibnamefont{Marshalek}},
  \bibinfo{journal}{Rev.\ Mod.\ Phys.} \textbf{\bibinfo{volume}{63}},
  \bibinfo{pages}{375} (\bibinfo{year}{1991}).


\bibitem[{\citenamefont{Thouless}(1961)}]{thou61}
\bibinfo{author}{\bibfnamefont{D.~J.} \bibnamefont{Thouless}},
  \emph{\bibinfo{title}{The Quantum Mechanics of Many-Body Systems}}
  (\bibinfo{publisher}{Academic}, \bibinfo{address}{London},
  \bibinfo{year}{1961}), \bibinfo{note}{p. 101}.

\bibitem[{\citenamefont{Brown}(1967)}]{bro67}
\bibinfo{author}{\bibfnamefont{G.~E.} \bibnamefont{Brown}},
  \emph{\bibinfo{title}{Unified Theory of Nuclear Models and Forces, second
  revised edition}} (\bibinfo{publisher}{North-Holland},
  \bibinfo{address}{Amsterdam}, \bibinfo{year}{1967}), \bibinfo{note}{chap. 7}.

\bibitem[{\citenamefont{Snyman}()}]{mythesis}
\bibinfo{author}{\bibfnamefont{I.}~\bibnamefont{Snyman}}, MSc thesis, University of Stellenbosch (2004);
  \eprint{nucl-th/0411059}.

\bibitem[{\citenamefont{Anderson}(1958)}]{and58}
\bibinfo{author}{\bibfnamefont{P.~W.} \bibnamefont{Anderson}},
  \bibinfo{journal}{Phys.\ Rev.} \textbf{\bibinfo{volume}{112}},
  \bibinfo{pages}{1900} (\bibinfo{year}{1958}).
  
\bibitem[{\citenamefont{Geyer et~al.}(1986)\citenamefont{Geyer,
  Engelbrecht, and Hahne}}]{geh86}
\bibinfo{author}{\bibfnamefont{H.~B.}~\bibnamefont{Geyer}},
  \bibinfo{author}{\bibfnamefont{C.~A.} \bibnamefont{Engelbrecht}},
  \bibnamefont{and} \bibinfo{author}{\bibfnamefont{F.~J.~W.} \bibnamefont{Hahne}},
  \bibinfo{journal}{Phys.\ Rev.\ C} \textbf{\bibinfo{volume}{33}},
  \bibinfo{pages}{1041} (\bibinfo{year}{1986}).
  
  
\bibitem[{\citenamefont{Dobaczewski et~al.}(1991)\citenamefont{Dobaczewski,
  Geyer, and Hahne}}]{dgh91}
\bibinfo{author}{\bibfnamefont{J.}~\bibnamefont{Dobaczewski}},
  \bibinfo{author}{\bibfnamefont{H.~B.} \bibnamefont{Geyer}},
  \bibnamefont{and} \bibinfo{author}{\bibfnamefont{F.~J.~W.} \bibnamefont{Hahne}},
  \bibinfo{journal}{Phys.\ Rev.\ C} \textbf{\bibinfo{volume}{44}},
  \bibinfo{pages}{1030} (\bibinfo{year}{1991}).
  

\bibitem[{\citenamefont{Brown et~al.}()\citenamefont{Brown, Gelman, and
  Kuo}}]{bgk04}
\bibinfo{author}{\bibfnamefont{G.~E.} \bibnamefont{Brown}},
  \bibinfo{author}{\bibfnamefont{B.~A.} \bibnamefont{Gelman}},
  \bibnamefont{and} \bibinfo{author}{\bibfnamefont{T.~T.~S.}
  \bibnamefont{Kuo}}, \eprint{nucl-th/0409071}.

\bibitem[{\citenamefont{Sakurai}(1994)}]{sak94}
\bibinfo{author}{\bibfnamefont{J.~J.} \bibnamefont{Sakurai}}
  (\bibinfo{collaboration}{edited by San Fu Tuan}),
  \emph{\bibinfo{title}{Modern Quantum Mechanics, revised edition}}
  (\bibinfo{publisher}{Addison-Wesley}, \bibinfo{address}{Reading, MA},
  \bibinfo{year}{1994}), \bibinfo{note}{p. 285}.

\end{thebibliography}
\end{document}